\documentclass[a4paper]{article}
\usepackage{graphicx}
\usepackage{amsmath}
\usepackage{multicol}
\usepackage[top=3cm, bottom=3cm, left=3cm, right=2cm]{geometry}

\makeatletter

\newenvironment{figurehere}
  {\def\@captype{figure}}
  {}
\makeatother

\begin{document}
\pagestyle{empty}

\title{
{\Large \bf 
Spectrometer up to 5~GeV
}
}
 
\author{
{\small 
Peter Maurer\footnote{Now at Harvard University}, \ Nicolas Delerue, \ David Urner, \ George Doucas 
}
}

%University of Oxford, OX1 3RH Oxford, United Kingdom
%Swiss Federal Institute of Technology ETH-Zürich, CH-8093 Zurich, Switzerland
%maurerpe@student.ethz.ch
%please contact to this person: nicolas.delerue@physics.ox.ac.uk

\date{
September 2007 
}

\maketitle
\thispagestyle{empty}

\begin{center}
{\bf 
To measure the energy spread of a energy electron beam up to $5\,GeV$ in a single shoot measurement the possible applications of an %%@
existing magnet as spectrometer has been discussed. For distinguishing the energy resolution and an optimal experimental setup %%@
numerical simulation, based on the measured magnetic field, have been performed. 
}
\end{center}

\begin{multicols}{2}{

\section{Problem} 

Electron pulses, which are aimed on to be studied, may have a comparative narrow energy speared (such as $\frac {\Delta E}{E}\, \sim %%@
\, 5\%$) \cite{Leemans} \cite{Nakamura}. On the other hand the electrons energy may vary between different electron bunches. Since %%@
the interest lies in knowing the energy distribution of a single bunch and not in the average distribution of the bunches it is %%@
required to measure the energy spread in a single shoot procedure. In addition the small spread in energy requires a high energy %%@
resolution. 

One possibility to determine the energy distribution in a single shoot measurement is given by using a dipole magnet as spectrometer. %%@
The discussed spectrometer can cover a large energy range ($\sim 100MeV$ to $5000MeV$) simultaneously. 

Since the desired magnet has been used for a different purpose (switching magnet) it is important to know its magnetic properties. %%@
Therefore a map of the magnetic field has been generated for different coil current. In addition the ramping characteristic of the %%@
magnet has been tested. 

To investigate the resolution the electron's trajectory has been studied by numerical simulation. These simulation allows to take the %%@
real magnetic field into account.

\section{Coordinate system} 

Depending on the problem it can be more convenient to deal with problems either in a lab's rest frame or in the beam's frame. 
In the lab's rest frame two different coordinate systems are used. The first is the cartesian system $(x,\,y,\,z)$), where the z axis %%@
is perpendicular to the pole surface, the x axis is along the connection between the source and the magnet's centre and the y is %%@
perpendicular to the x and z axis (Fig. \ref{fig:coord_system}). 

%\begin{figure}[htbp]
\begin{figurehere}
\begin{center}
\includegraphics[width=7cm]{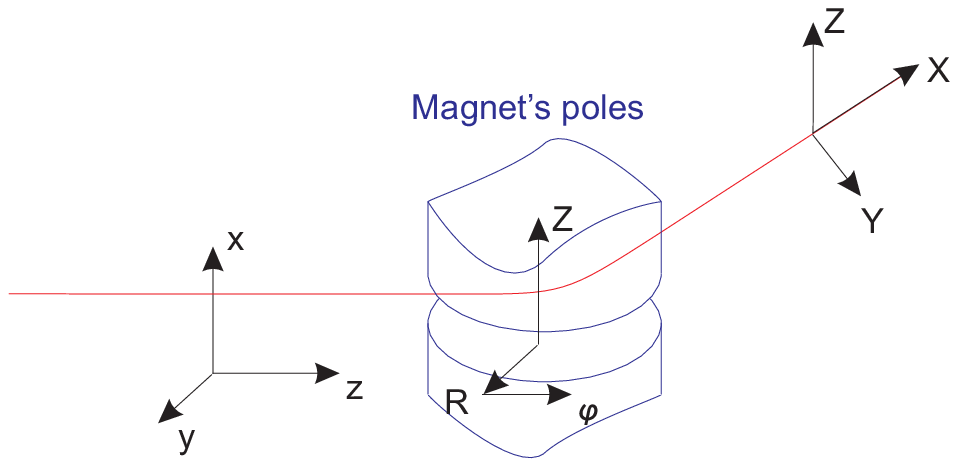}
\caption{ Sketch of the orientation for the different coordinate systems. The magnet's pole are pictured blue. The beam's trajectory %%@
is red and the three coordinate systems are black. 
\label{fig:coord_system} }
\end{center}
\end{figurehere}

The coordinate system $(X,\,Y,\,Z)$ is parametrised so that the Z axis is along the propagation of the beam's centre, the Y axis is %%@
perpendicular to Z and z. And the X axis is perpendicular to Y and Z (Fig. \ref{fig:coord_system}). 
Due to the radial symmetry of the magnetic field it is some time more convenient to use a cylindrical coordinate system %%@
$(r,\,\varphi,\,Z)$. 

\section{Discussion on the Magnet} 

\subsection{Magnet's physical data} 

The studied magnet was previously used as switching magnet. It was fabricated in 1989 by Danfysik (serial number 89374). It generates %%@
a field of $1.15\,T$ at $87\,A$. The physical radius of the magnet is $\sim256\,mm$ the magnetic radius is $\sim(167.0\pm 0.6)\,mm$ %%@
(see Fig. \ref{fig:magnet}). 

\begin{figurehere}
\begin{center}
\includegraphics[width=7cm]{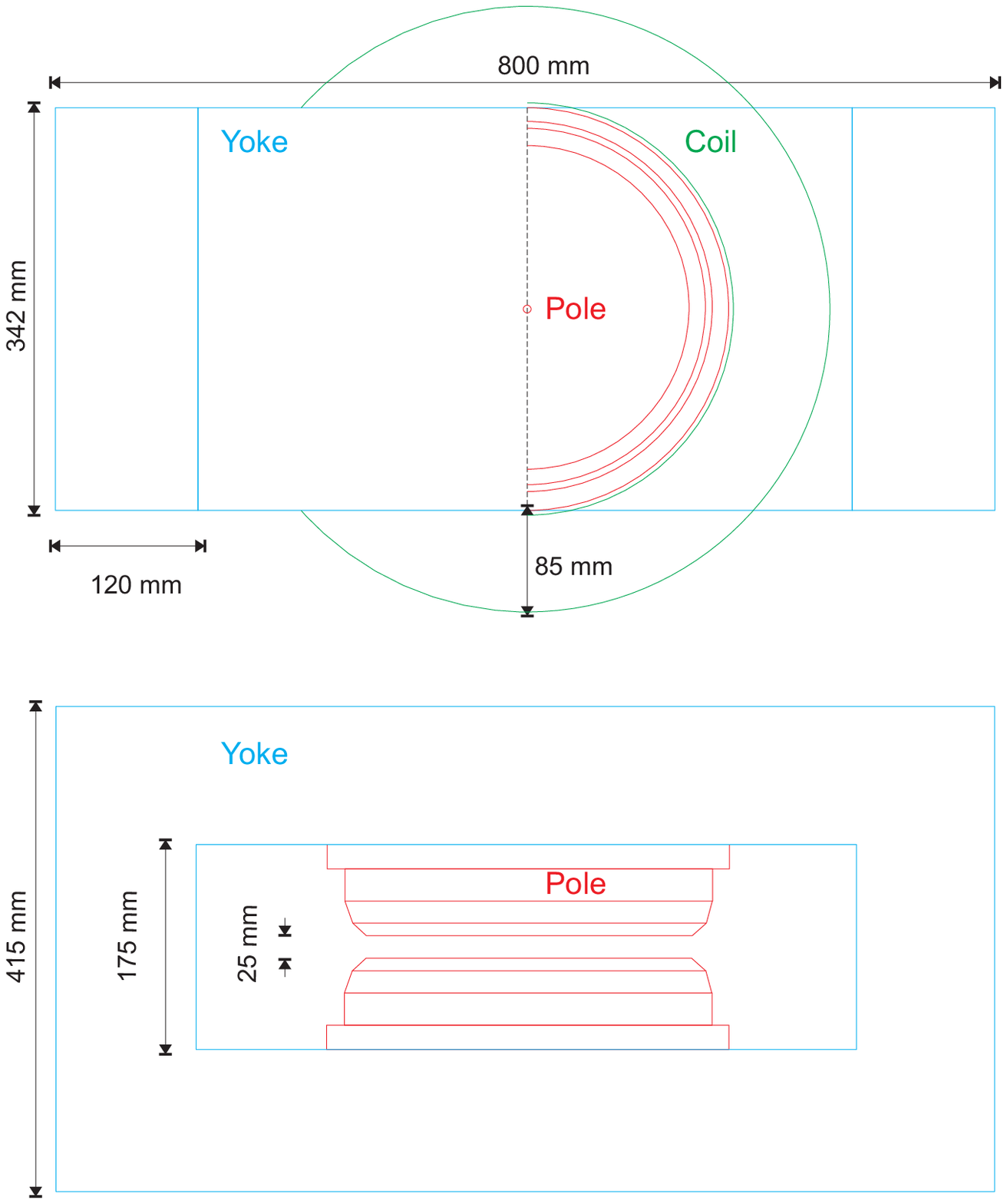}
\caption{ Sketch of the magnet which will be used as the spectrometer. The first picture shows the ground plan and the second the %%@
sheer plane of the magnet. The yoke is blue, the magnet poles are red and the coil is green pictured. In the second drawing the coil %%@
is omitted. 
\label{fig:magnet} }
\end{center}
\end{figurehere}

A new vacuum chamber which allows to cover a angular spread of $\pm45^\circ$ or $-10^\circ$ to $80^\circ$ is being designed.

%\begin{figure}[h!]
%\begin{center}
%\includegraphics[height=6cm]{Vacuum_chamber.eps}
%\caption{ 
% 
%\label{fig:Vacuum_chamber} }
%\end{center}
%\end{figure}

\subsection{Measurement methods} 

To check the field homogeneity the magnetic field of the spectrometer was mapped for different coil currents and different positions. %%@
To measure the field's componentes a transversal and a longitudinal hall probe ($\pm$ 1mT) were used. 

\begin{figurehere}
\begin{center}
\includegraphics[width=7cm]{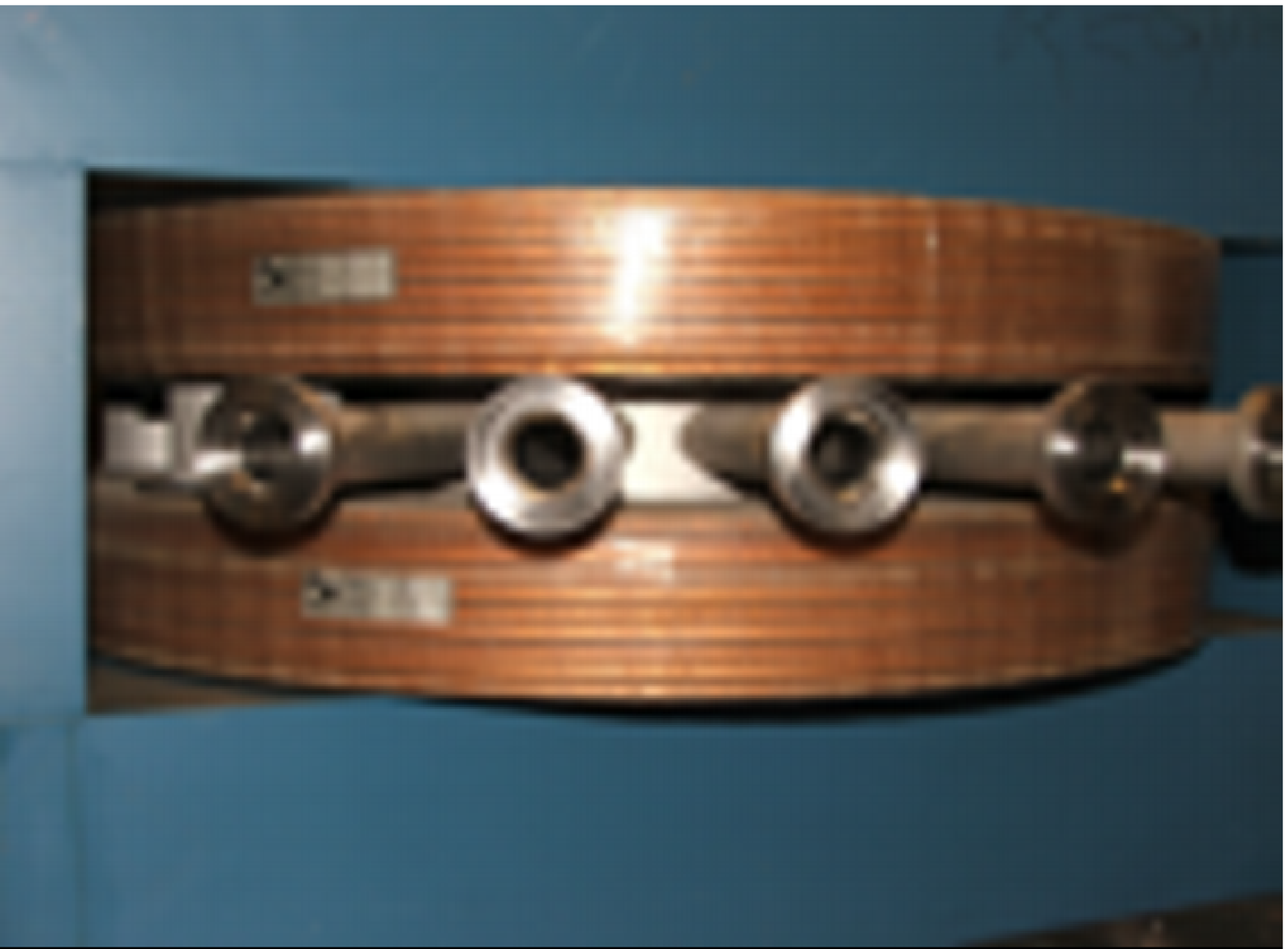}
\caption{ Photo of the magnet with the old vacuum chamber. To measure the $B_z$ field the hall probe was inserted in the different %%@
tubes. 
\label{fig:vacuum_tube_old} }
\end{center}
\end{figurehere}

Since the largest part of the field is contributed by the component along z a good knowledge of $B_z$ is crucial for determining the %%@
beam's deflection angle. Therefore a profile of the vertical field was measured as a function of the radius for %%@
$-30^o,\,-15^o,\,0^o,\,15^o,\,30^o$ and $45^o$ deflection angle. These angles were given by the alignment of the previous vacuum tube %%@
(Fig.~\ref{fig:vacuum_tube_old}), which had ports at $0^o,\,\pm 15^o,\,\pm 30^o$ and $\pm 45^o$. To insert the hall probe into the 

different vacuum tubes the probe was fixed on a pipe of stainless steel which fitted in the vacuum tubes. The position of the hall %%@
probe was controlled by measuring the probe's position relative to the vacuum tube's entrance ($\pm$ 0.5mm). The twisting angle %%@
$\beta$ between the hall probe and the field was maximally $0.5^\circ$. 
If the B field is assumed to be radial symmetric, it can be written as $B_z(r,B_{max})\,=\,B_{max}\,f(r)$. The true field is then %%@
simply $B\,=\,B_z(r,B_{max})\,cos(\beta)$. The error is then given by 

$ $

$$\begin{array}{c}
\Delta B\,= (cos^2\beta\,\left(\frac{\partial B_z}{\partial r}\,\Delta r \right)^2+ \\ +cos^2\beta\,\left(\frac{\partial %%@
B_z}{\partial B_{max}}\,\Delta B \right)^2+B_z^2\,sin^2\beta\,(\Delta \beta)^2)^{\frac{1}{2}}
\end{array}$$

$ $

Using $\frac{\partial B_z}{\partial B_{max}}\,=\,f(r)$ and $\frac {sin^2\beta}{cos^2\beta}\,\approx \beta^2$ the relative error is %%@
than given by 

$ $

$$\frac {\Delta B}{B}\,=\,\sqrt{\left(\frac{\partial f(r) / \partial r}{f(r)}\,\Delta r \right)^2+\left(\frac{\Delta B}{B_{max}} \right)^2+\beta^2\,(\Delta \beta)^2 }$$

$ $

For the experiment this becomes 

$ $

$$
\frac {\Delta B}{B}\,=\,\left(\left(\frac{\partial f(r) / \partial r}{f(r)}\,0.5\,mm \right)^2 +\\ +\left(\frac{1\,mT}{B_{max}} \right)^2+\beta^2\,(0.009)^2 \right)^{\frac{1}{2}}\,
$$

$$\approx\,\sqrt{\left(\frac{\partial f(r) / \partial r}{f(r)}\,0.5\,mm \right)^2+\left(\frac{1\,mT}{B_{max}} \right)^2}$$

$ $

Where $\frac{\partial f(r) / \partial r}{f(r)}$ takes value between 0 and $0.03\,mm^{-1}$. Therefore the uncertainty depends strongly %%@
on the position in the field. At the magnet edge the relative error is up to $\frac {\Delta B}{B}\,\approx\,15\%$. For the magnet's %%@
bulk it is on the other hand in the range of $\frac {\Delta B}{B}\,\approx\,1^o/_{oo}$. 

$ $

Since the magnet's poles are quite parallel and have a large diameter $(\sim 170\,mm)$ compared to the pol gap $(\sim 25\,mm)$ it %%@
likely that the field in the magnet's bulk does not vary much in z and has only small radial component. Therefore it is more %%@
important to map the field at the magnet's edge. For simplicity the field has only been measured in the (r,z) plane. The stray field %%@
measurements have been done after removing the original vacuum chamber. To determine the exact position of the hall probe the probe %%@
was fixed to a table which was able to be adjusted in height with $\pm 1\,mm$ accuracy. The radial position was controlled by a %%@
measuring stick which was glued to the table ($\pm 1\,mm$). The experimental setup is given in Fig. \ref{fig:r_Z_Messung}. The radial %%@
field was measured by using a longitudinal and the vertical field by using a transversal hall probe. 

\begin{figurehere}
\begin{center}
\includegraphics[width=7cm]{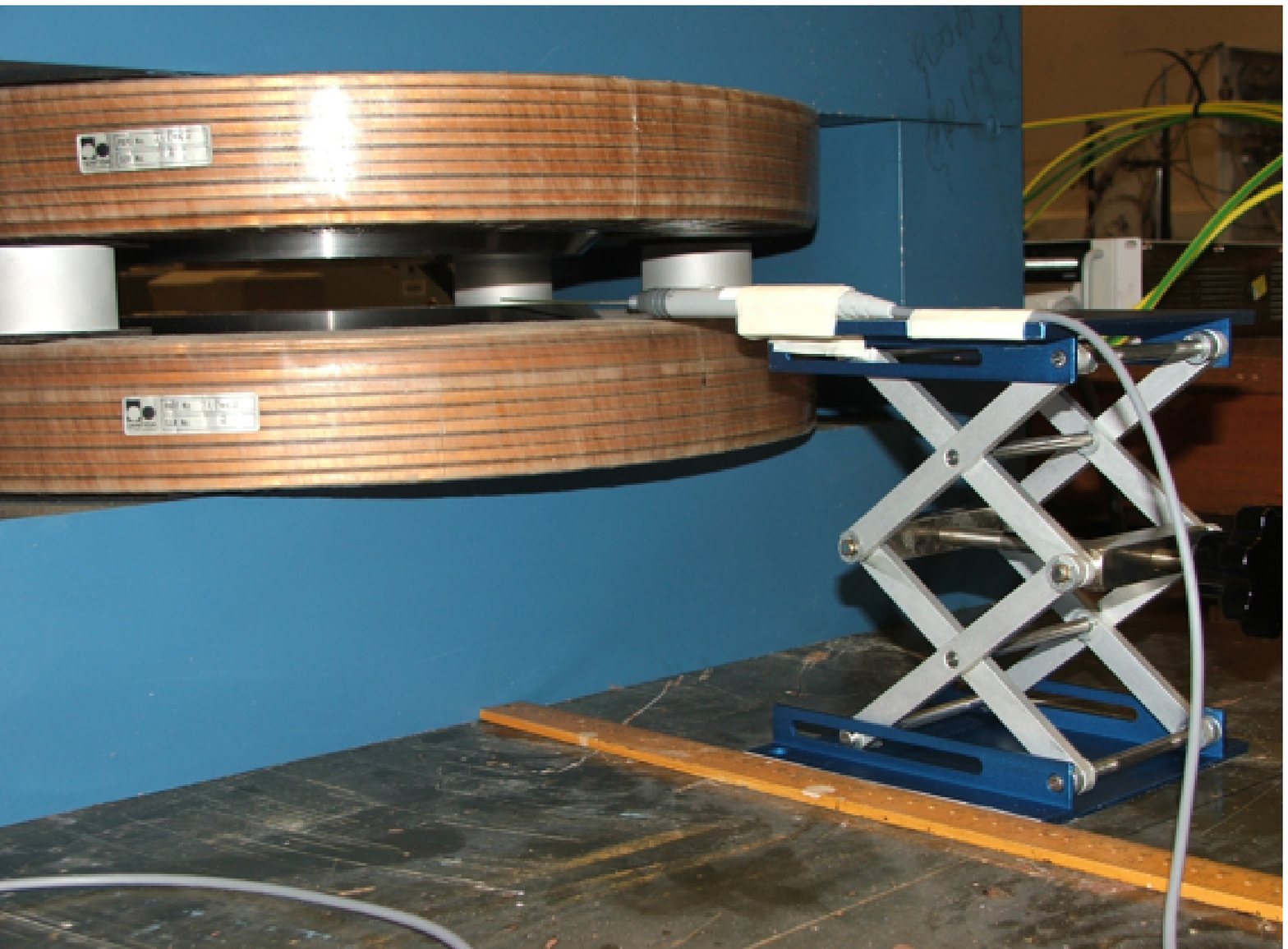}
\caption{ Experimental setup to map the $B_z\,ans\,B_r$ component of the magnetic stray field as a function of r and z. The %%@
uncertainty in r and z is $1\,mm$. 
\label{fig:r_Z_Messung} }
\end{center}
\end{figurehere}

Since the radial field can be very small and the vertical field is large $(\approx 1.150\,T)$ a small off zero angle between the z %%@
axis and the hall probes surface gives a flux that can not be neglected. Therefore the measured radial field has to be corrected by %%@
this flux contributed from the vertical field. The contribution from the vertical field was determined by inserting the probe inside %%@
the magnet's bulk where the radial field should be zero. This allows to calculate a simple correction term. 

\subsection{Results on Magnetic properties} 

The magnetic field perpendicular to the poles $B_z$ has been measured as a function of the radial position Fig.\ref{fig:field}. If %%@
the magnetic field is normalized by $f=\frac{B_z-B_{earth}}{B_{max}}$, it can be seen (Fig.\ref{fig:field_norm}) that the field's %%@
normalized shape does not change with the coil current. 

\begin{figurehere}
\begin{center}
\includegraphics[width=7cm]{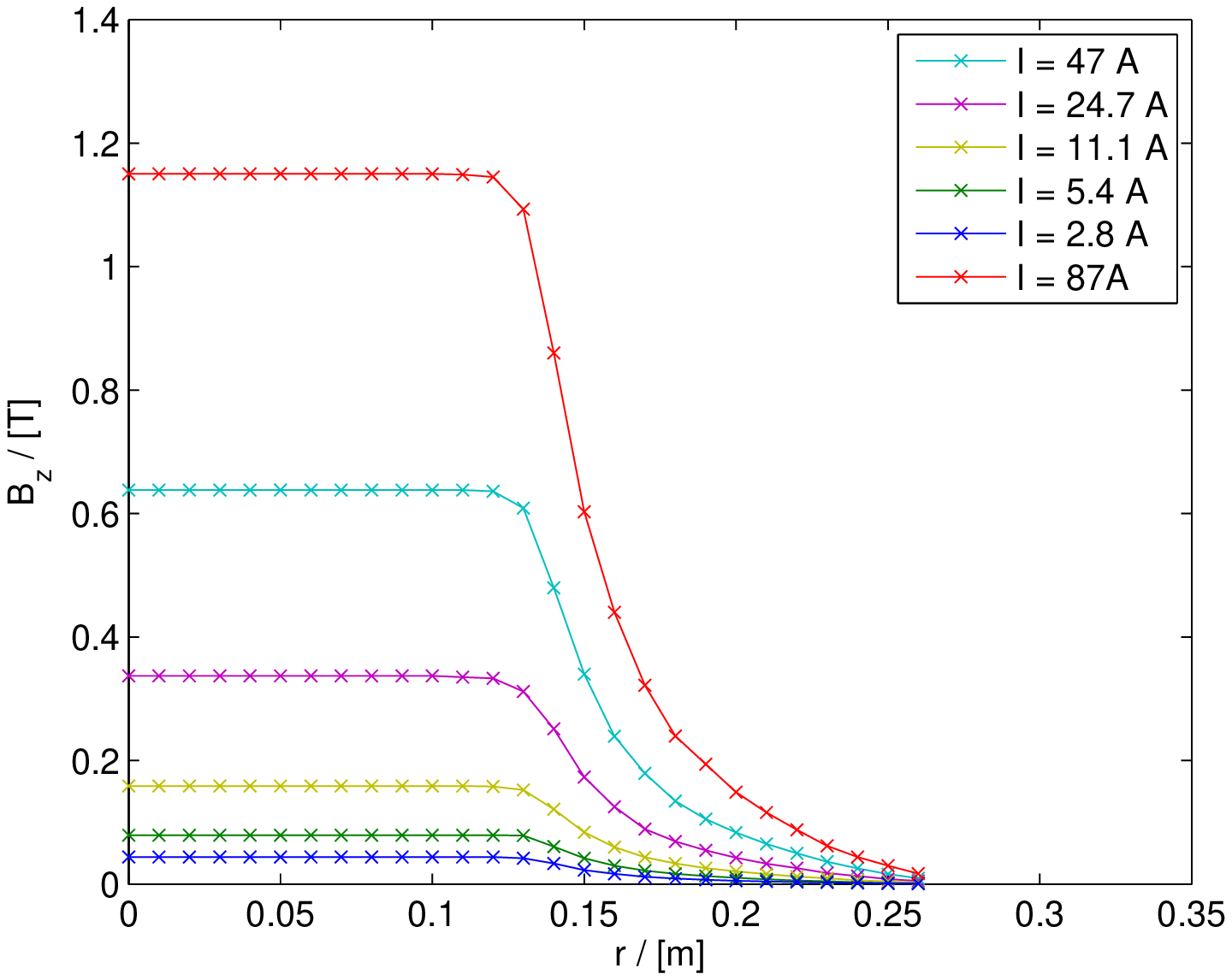}
\caption{ The magnetic field as distance from the magnet's centre. The measurements were done along the connection line between 
port 4 and 8 %(see Fig. \ref{field_phi_bild}).
\label{fig:field} }
\end{center}
\end{figurehere}

\begin{figurehere}
\begin{center}
\includegraphics[width=7cm]{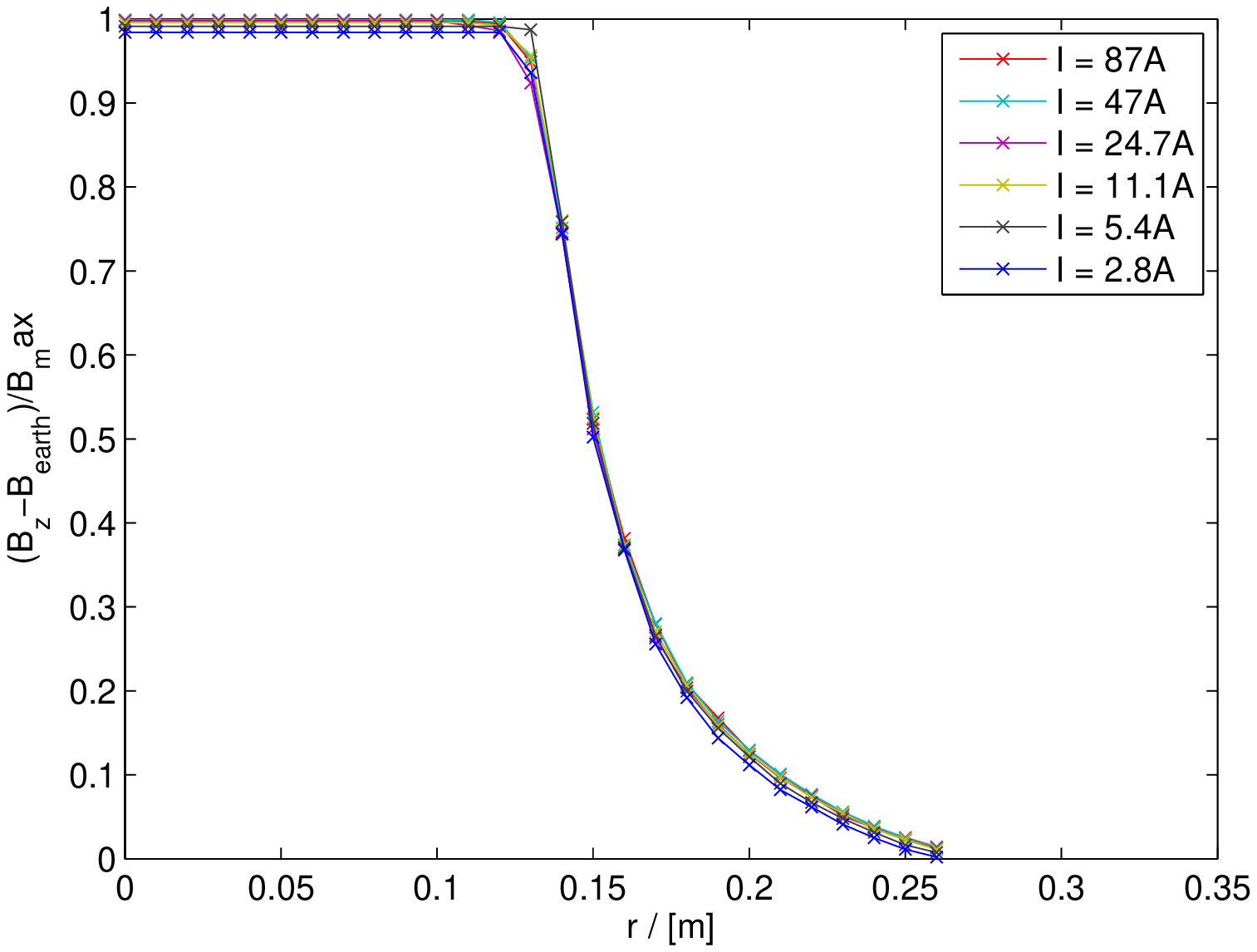}
\caption{ The results from %Fig.\ref{fig:field} normalized after subtracting the earth field. It shows that the field's envelope %%@
keeps its shape even for low coil current. 
\label{fig:field_norm} }
\end{center}
\end{figurehere}

This allows to introduce the magnetic radius as 

$ $

$$R_{magnetic}\,=\,\int^{\infty}_0 f(x)\: \mathrm{d}x$$

$ $

with $B_{max}$ as the maximal field for a certain coil current. Due to the measurement the magnetic radius 
$R_{macnetic}\,=\,(167.0\pm 0.6)\,mm$ can be found. 

The length of the stray field $\Delta R_{stray}$ defined as the region where the field drops from $95\%$ to $5\%$ was determined as %%@
$\Delta R_{stray}\,=\,(110 \pm 5) \, mm$. The ratio between the stray field length and the magnetic radius was found to be $\frac %%@
{\Delta R_{stray}}{R_{magnetic}}\,=\,0.66 \pm 0.02$

To make sure that the field is radial symmetric the measurement from Fig. \ref{fig:field} has been done for different directions.The %%@
radial symmetry can clearly be seen in Fig. \ref{fig:field_phi_bild}.

\begin{figurehere}
\begin{center}
\includegraphics[width=7cm]{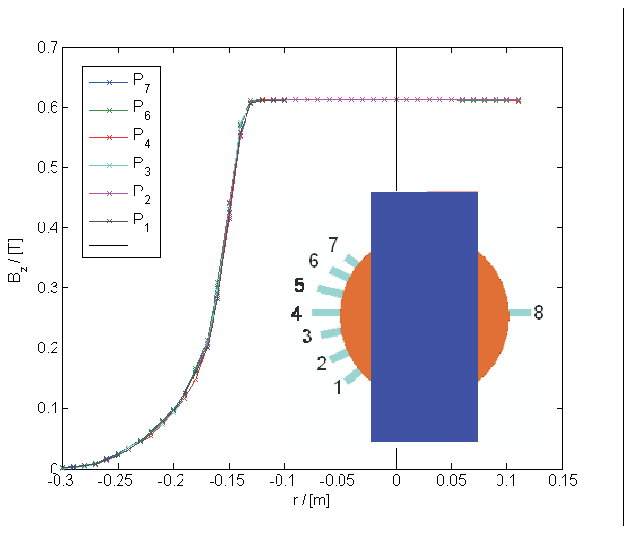}
\caption{ The magnetic field as a function of the distance to the magnet's centre for different directions as showed in the inserted %%@
sketch. The measurement have been done for $47\,A$ coil current. 
\label{fig:field_phi_bild} }
\end{center}
\end{figurehere}

Far from the edge inside the magnet the field is parallel to the z axis and far outside the field is negligible. Therefore the field %%@
has only been measured at the magnet's edge as a function of its radial and vertical position. The results can be seen in Fig. %%@
\ref{fig:Field_line}. The magnetic field lines are quite parallel. The measured radial field is smaller than $15\%$ of the vertical %%@
field. As it will be seen in the part on simulation corrections due to the vertical position and the radial field are small. 

\begin{figurehere}%[htbp]
\begin{center}
\includegraphics[width=7cm]{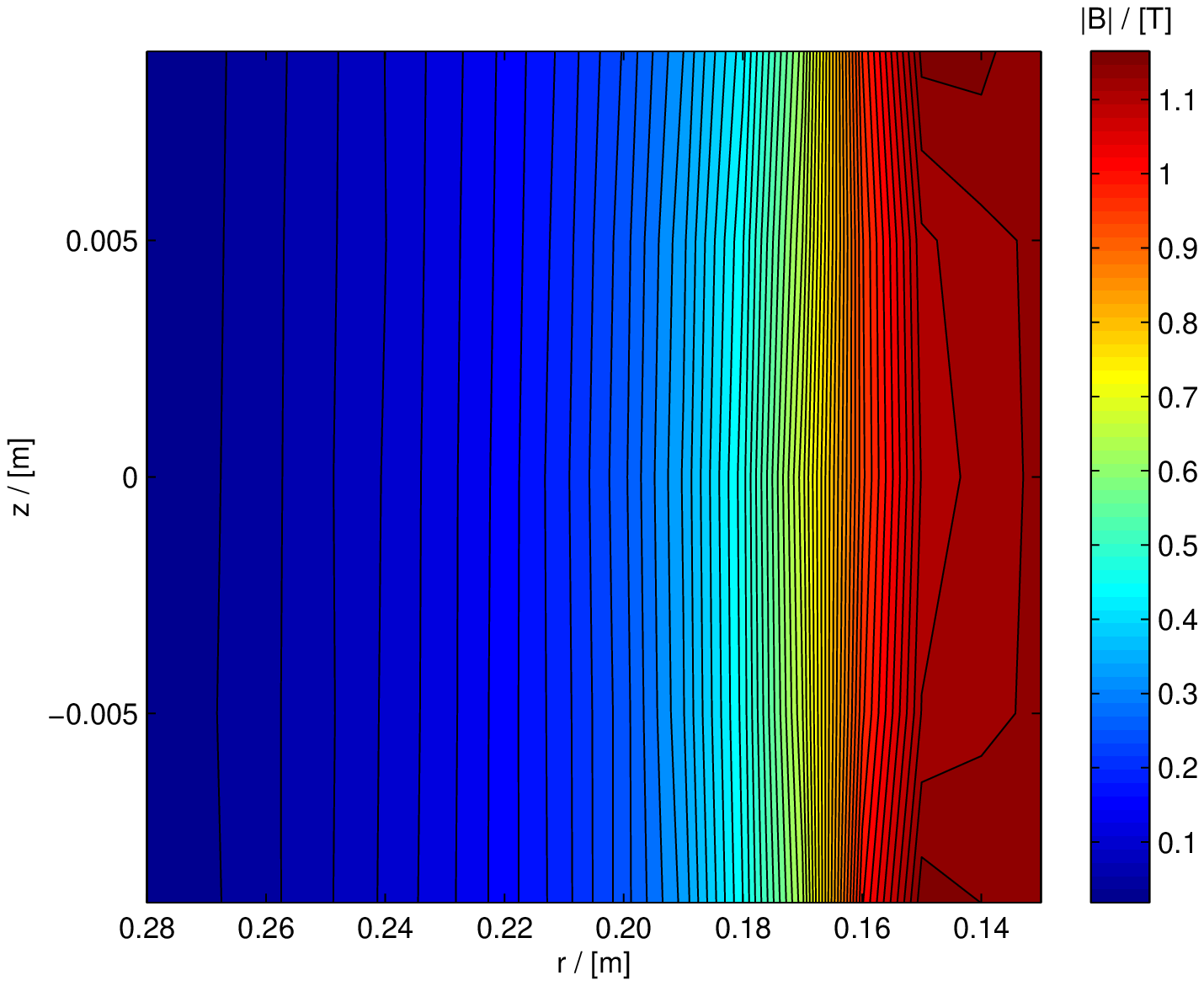}
\caption{ The absolute value of the magnetic field as a function of the radial and the vertical position. The vertical position %%@
$z\,=\,0$ is the midpoint of the pol gap. The measurement have been done at 87 A coil current. 
\label{fig:Field_line} }
\end{center}
\end{figurehere}

The magnetic field for the spectrometer was simulated in Poisson$/$Superfish \cite{Superfish}. The program Poisson$/$Superfish solves %%@
the two dimensional Maxwell equation for certain boundary conditions. The boundary conditions were in such a way chosen that they %%@
represent a (x,y) cut though the magnet's centre.  The resulting field lines are given in \ref{fig:Superfish}. Since the coil %%@
consists of 140 windings a total current of $12180\,A\,=\,140\,\cdot\,87\,A$ was used for the calculations. A maximal field of $1.15 %%@
T$ could be found. This is in good agrement with the $(1.150 \pm 1)\,T$, which where found in the measurements. 

\begin{figurehere}
\begin{center}
\includegraphics[width=5cm,angle=-90]{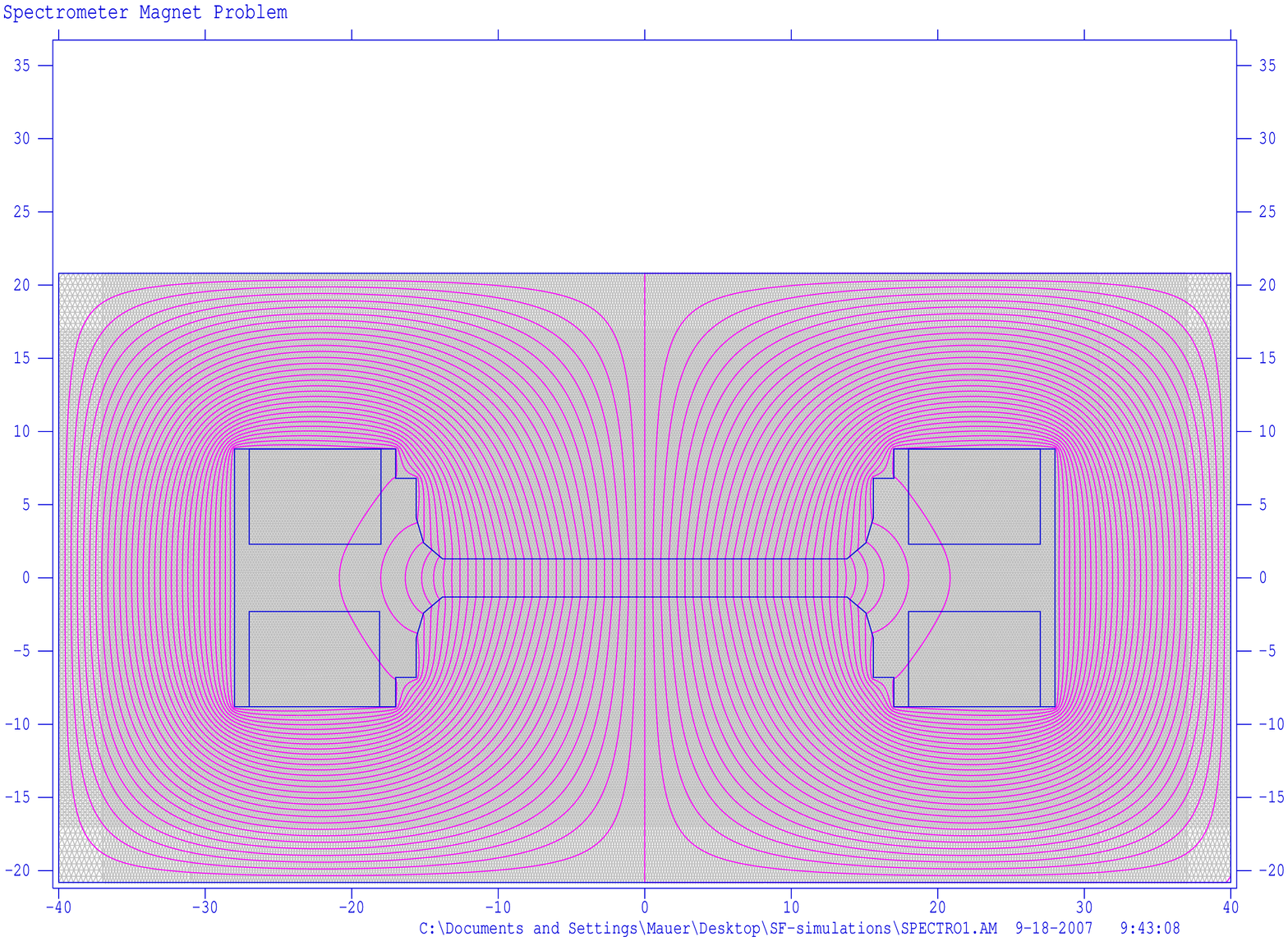}
\caption{ Simulated field for the two dimensional Maxwell equations, with the boundary conditions given by a x-y cut through the zero %%@
point. 
\label{fig:Superfish} }
\end{center}
\end{figurehere}

To determine the ramping characteristic and check on memory effects the magnetic field was changed adiabaticly. To make sure that %%@
changes were adiabatic the coil current was slowly $(5 A \, min^{-1})$ shifted. Two cycle starting at $0 A$, going to $47 A$, $87 A$ %%@
respectively and back to $0 A$ were measured. The magnetic field for the larger cycle is given in Fig. \ref{fig:Zyklus_Hyst}. It can %%@
be found that 

$$B(I_{\gamma _1}(t_{end}))<B(I_{\gamma _2}(t_{end}))$$

with $I_{\gamma _1}(t) < I_{\gamma _2}(t)\,\forall \, t<t_{end}$ and $I_{\gamma _1}(t_{end}) = I_{\gamma _2}(t_{end})$. $\gamma$ is the %%@
path that was used to change the current. The differences $B(I_{\gamma _2}(t_{end}))-B(I_{\gamma _1}(t_{end}))$ are shown in Fig. %%@
\ref{fig:Zyklus_Hyst_diff}. 

\begin{figurehere}
\begin{center}
\includegraphics[width=7cm]{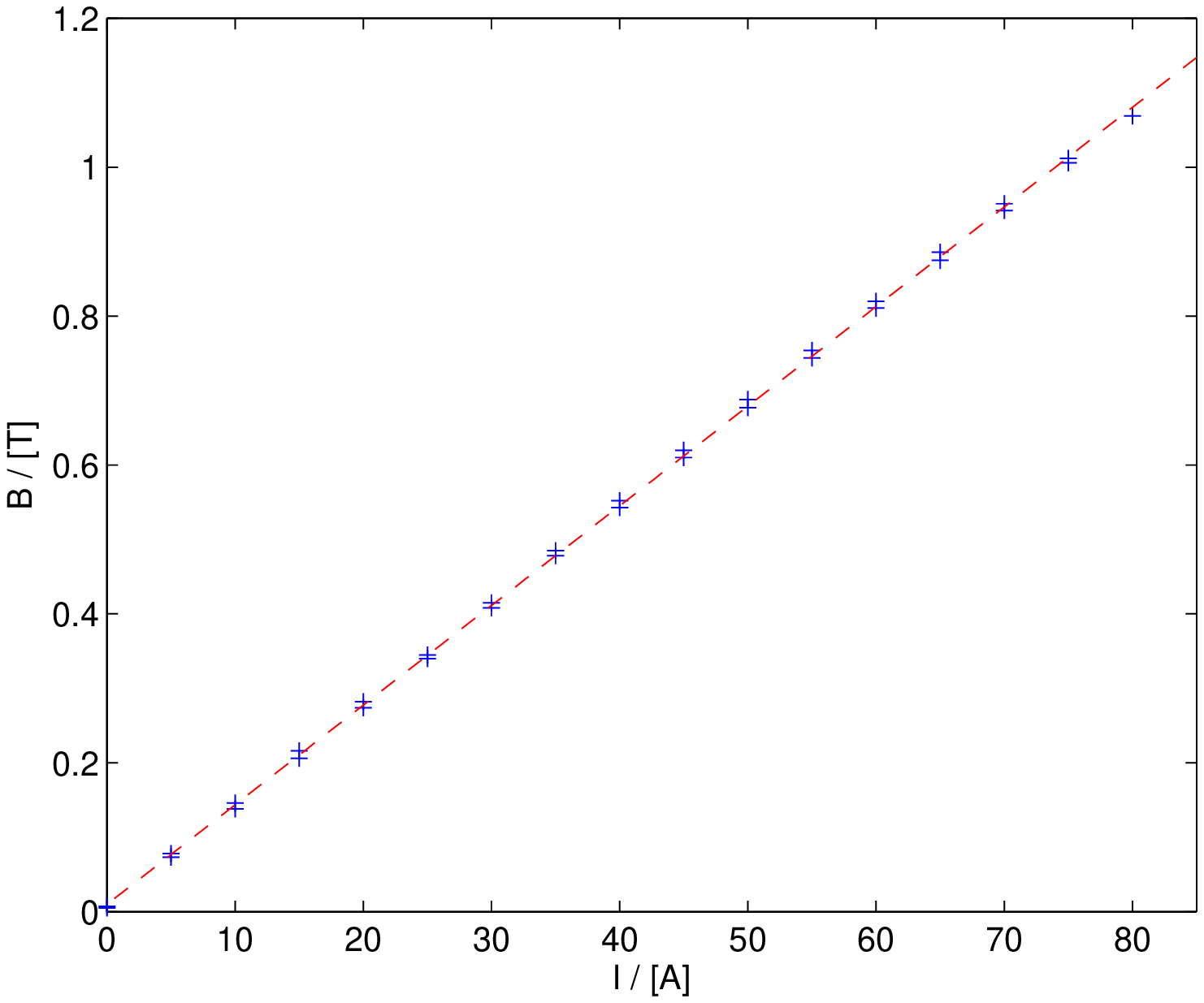}
\caption{ The plot shows the magnetic field in the magnets centre vs. the coil current ($0\,A\,to\,87\,A$). It can bee seen that the %%@
field is nearly proportional to the current. 
\label{fig:Zyklus_Hyst} }
\end{center}
\end{figurehere}

\begin{figurehere}
\begin{center}
\includegraphics[width=7cm]{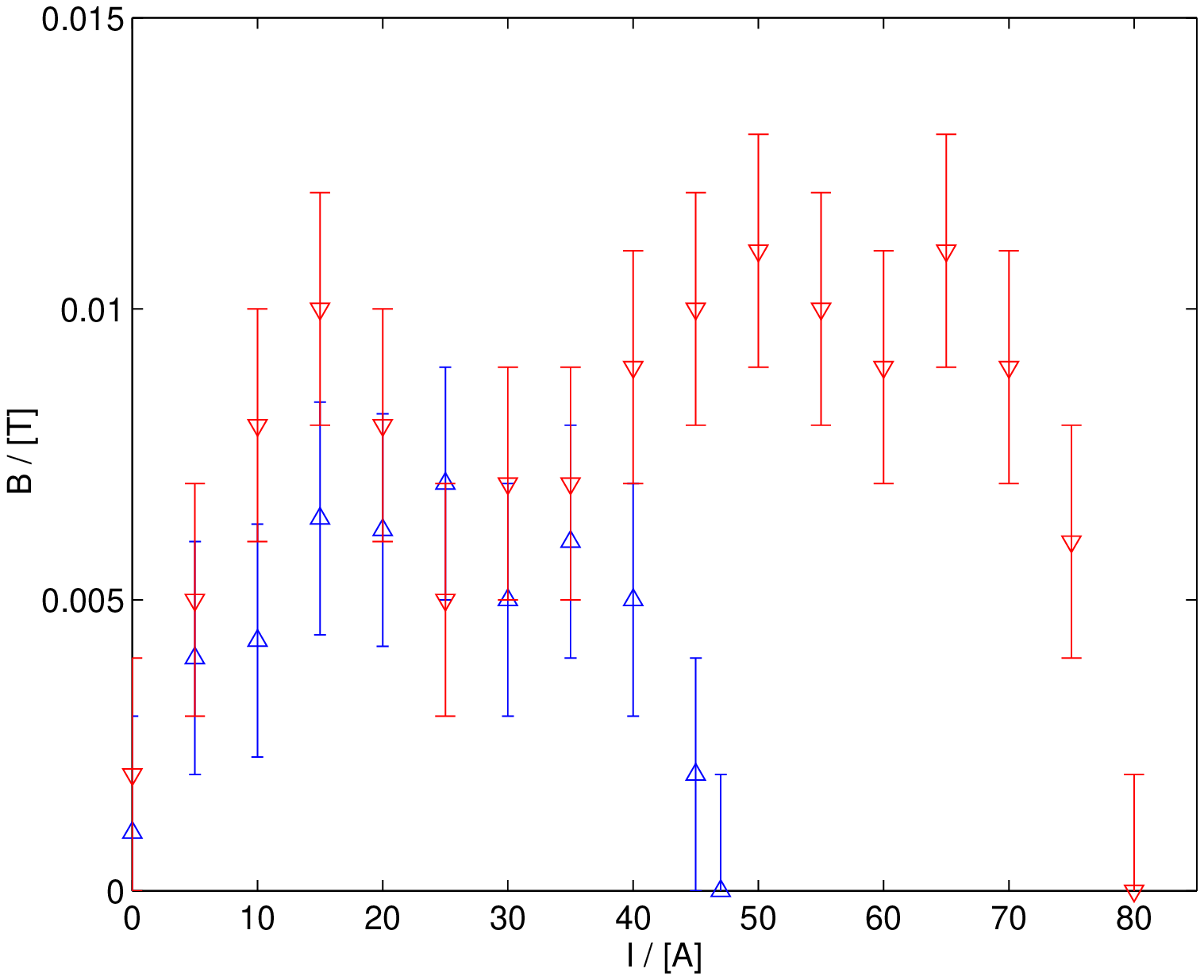}
\caption{ The plot shows the difference for the $B(I_{\gamma _2}(t_{end}))-B(I_{\gamma _1}(t_{end}))$ as a function of the coil %%@
current. The current was changed slowly ($5\,A\,min^{-1}$) to stay in the adiabatic limit. 
\label{fig:Zyklus_Hyst_diff} }
\end{center}
\end{figurehere}

\section{Discussion on Simulation} 

\subsection{Simulation methods} 

To determine the trajectory of an electron in the spectrometer it is required to take the real field of a magnet into account. It was %%@
assumed that the magnetic field is constant for each iteration step. Using the relativistic momentum %%@
$\vec{p}\,=\,m_0\,\gamma\,\vec{v}$ the equation of motion is given by 

$ $

$$\frac{d\vec{p}}{dt}\,=\,q\,\vec{v}\,\wedge\,\vec{B}$$

$ $

$$m_0\,\beta\,\gamma\,\left(\gamma^2\,\beta^2+1\right)\,\frac{d\vec{v}}{dt}\,=\,q\,|\vec{v}|\,\vec{n}_v\,\wedge \vec{B}$$

$ $

Since there is only a Lorentz force acting on the electron $\gamma$ and $\beta$ remain constant. Using this the differential equation %%@
simplifies to 

$ $

$$m_0\,\gamma\,\left(\gamma^2\,\beta^2+1\right)\,\beta^2\,c^2\,\frac{d^2\vec{x}}{ds^2}\,=\,q\,\beta\,c\,\vec{n}_v\,\wedge\,\vec{B}$$

$ $

Where the chain rule has been applied to write the problem as a function of the trajectory length s. 

With this assumption the equation of motion can be solved for each iteration step. To get an idea of the algorithm's robustness an %%@
electron's trajectory was simulated in a homogeneous magnetic field. Since an analytic expression for an electron's motion in such a %%@
field exists ($\rho\,=\,\frac{p}{e\,B}$) it is possible to compare the derivation between the analytical and numerical solution Fig. %%@
\ref{fig:Fehler_Numeric}. It can be seen that for reasonable step size the true value of $\rho$ is achieved with a high accuracy %%@
($\approx\,1^o/_{oo}$). 

\begin{figurehere}
\begin{center}
\includegraphics[width=7cm]{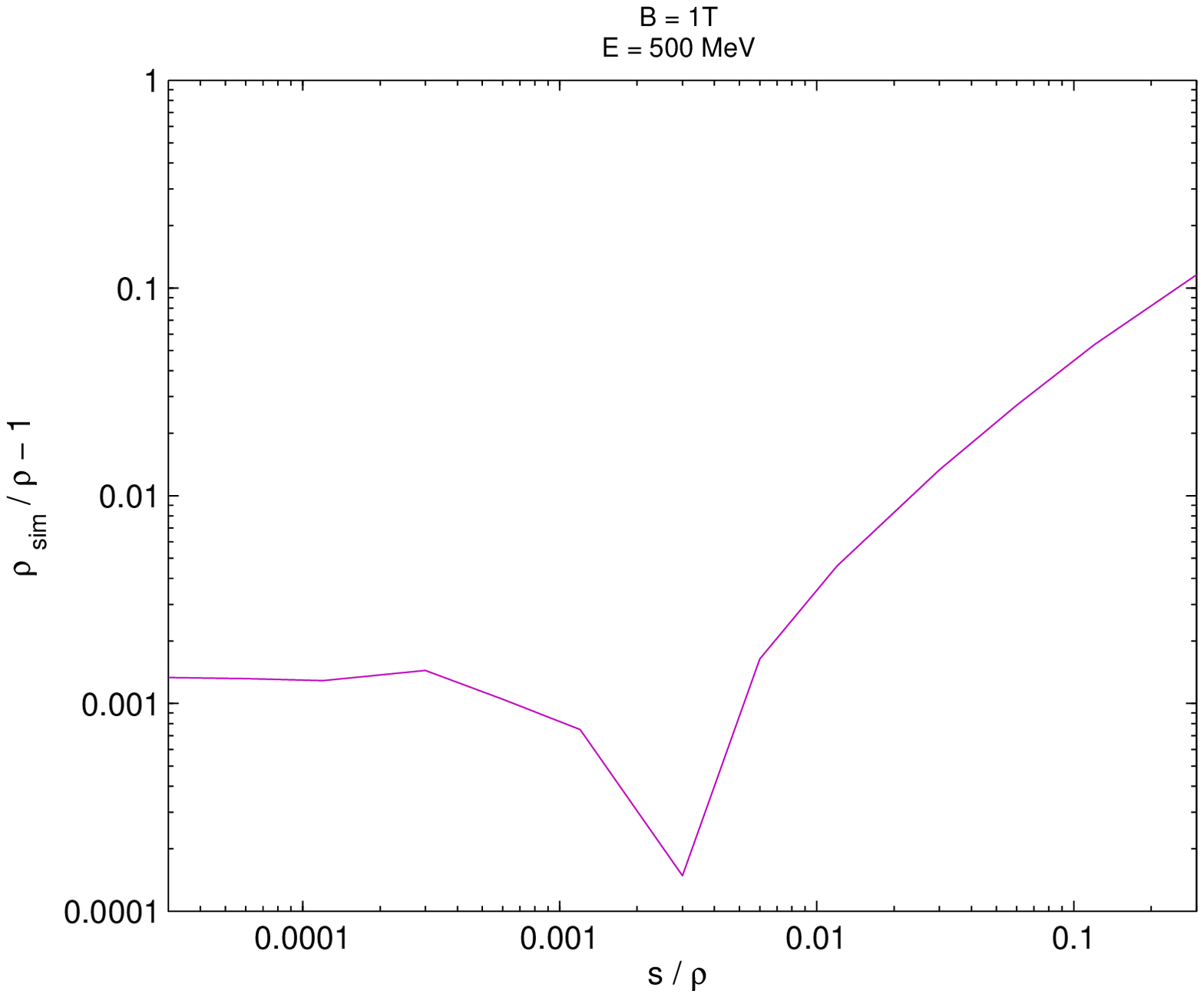}
\caption{ To estimate the simulation's accuracy the trajectory of an electron in a homogeneous field has been calculated and compared %%@
with the theoretical value. Where s is the step size, $\rho$ is the theoretical bending radius and $\rho _{sim}$ is the simulated. 
\label{fig:Fehler_Numeric} }
\end{center}
\end{figurehere}

For all the simulation an experimental setup as shown in Fig.\ref{fig:setup} was used. In the simulation either the magnetic field %%@
measured by Danfysik or the one measured by us were used. 

\begin{figurehere}
\begin{center}
\includegraphics[width=7cm]{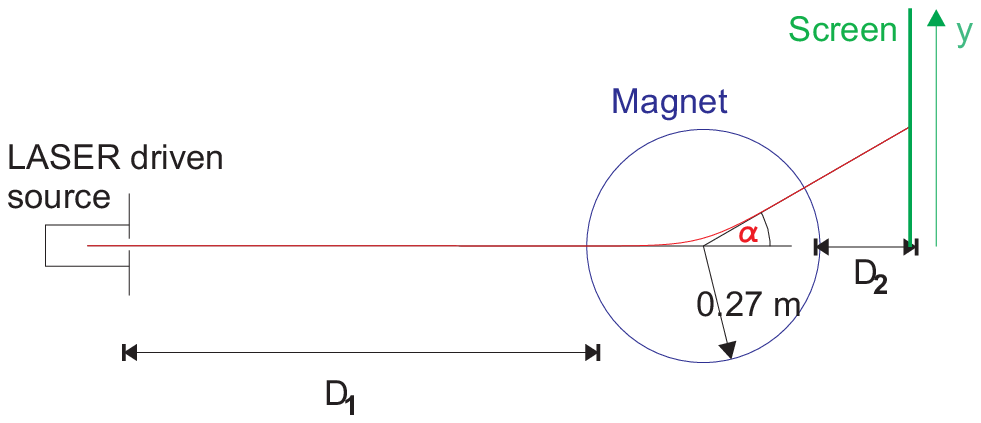}
\caption{ Schematically picture of the setup used for the simulation. This setup have been chosen for the all the simulations. The %%@
detector is in a cylindrical shape placed behind the spectrometer. The magnet's physical diameter is 540mm the magnetic diameter is %%@
330mm. 
\label{fig:setup} }
\end{center}
\end{figurehere}

\subsection{Results from simulation} 

For the measured magnetic field the deflection of an electron beam can be simulated as a function of its energy as seen in Fig. %%@
\ref{fig:Bild_von_traject}. If this trajectories are compared with the trajectories for a field given by %%@
$\vec{B}\,=\,B_{max}\,\vec{e}_z\,\chi _{\sqrt{x^2+y^2}\leq R_{magnetic}}$, an error in the deflection angle of maximally $2\%$ can be %%@
found Fig. \ref{fig:anal_vs_simul}. Since an error of $2\%$ is to large the problem can not be handled analytically (as in the appendix) %%@
but has to be treated numerically. All the further discussions will base on numerical simulation. 

\begin{figurehere}
\begin{center}
\includegraphics[width=8cm]{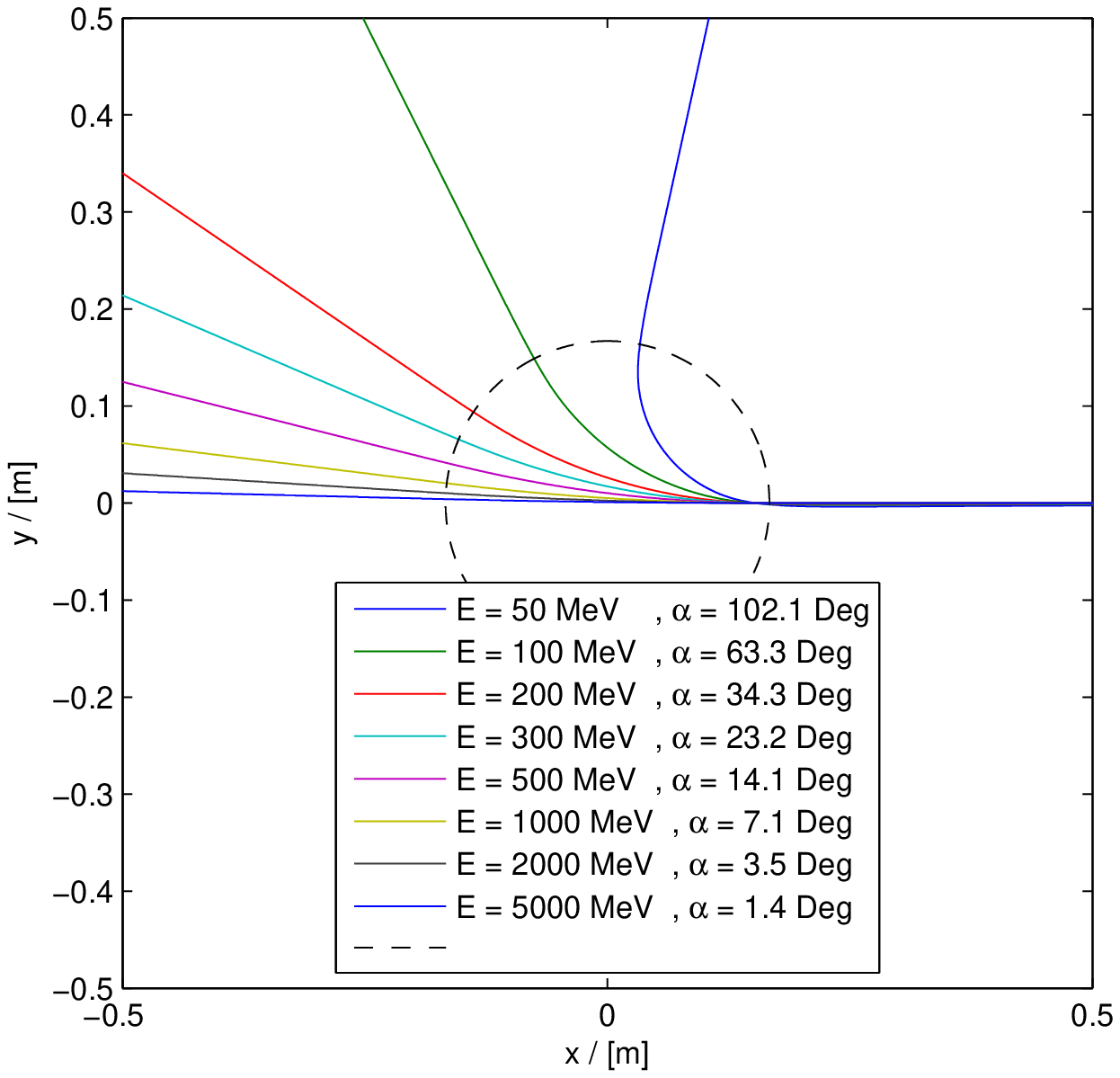}
\caption{ The trajectories for electron beams at different energies in the real magnetic field ($B \sim 1.15\,T$). The broken circle %%@
indicates the spectrometer's magnetic radius. 
\label{fig:Bild_von_traject} }
\end{center}
\end{figurehere}

\begin{figurehere}
\begin{center}
\includegraphics[width=7cm]{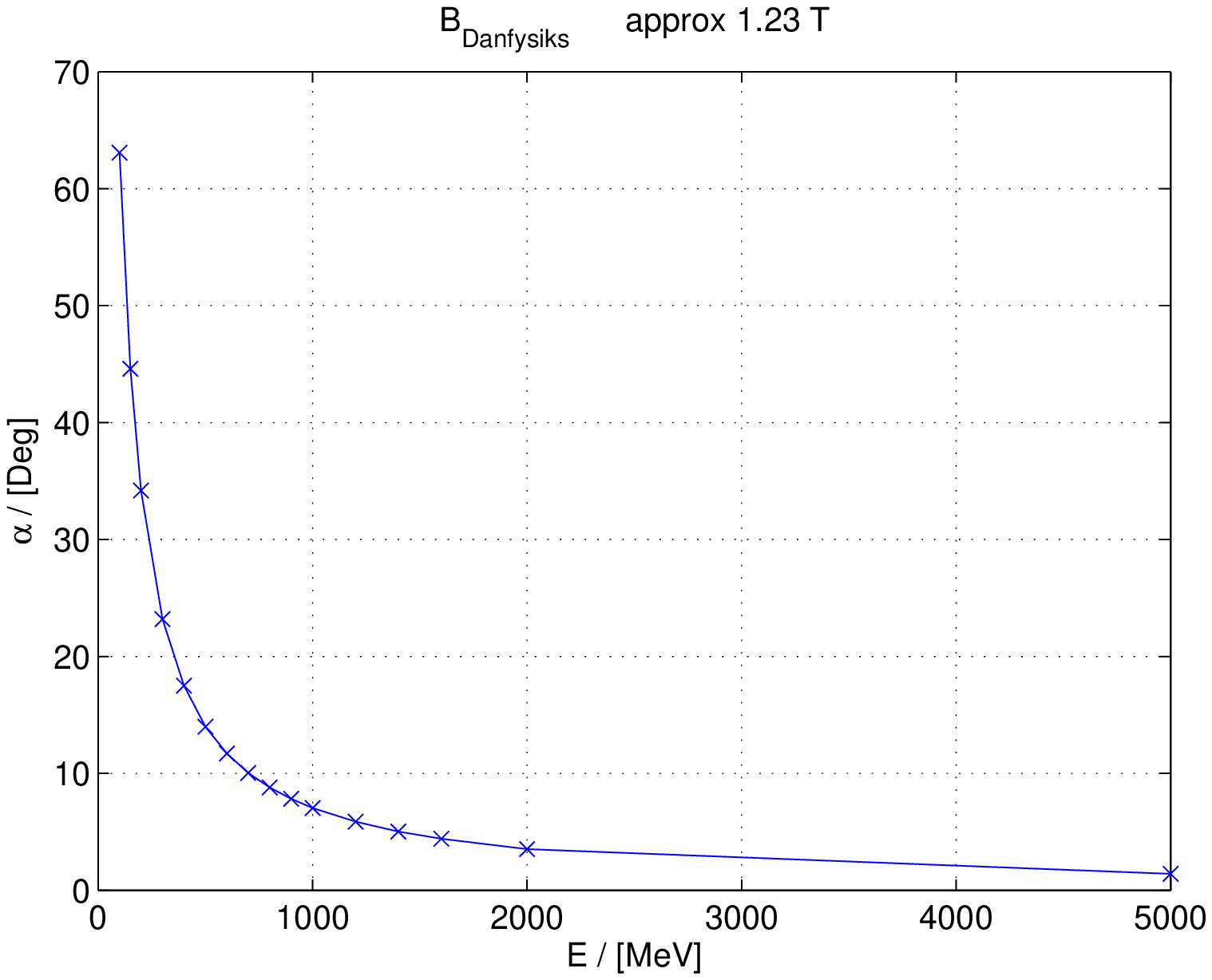}
\caption{ The plot shows the energy $E$ vs. the bending angle $\alpha$. 
\label{fig:bending_energy} }
\end{center}
\end{figurehere}

\begin{figurehere}
\begin{center}
\includegraphics[width=7cm]{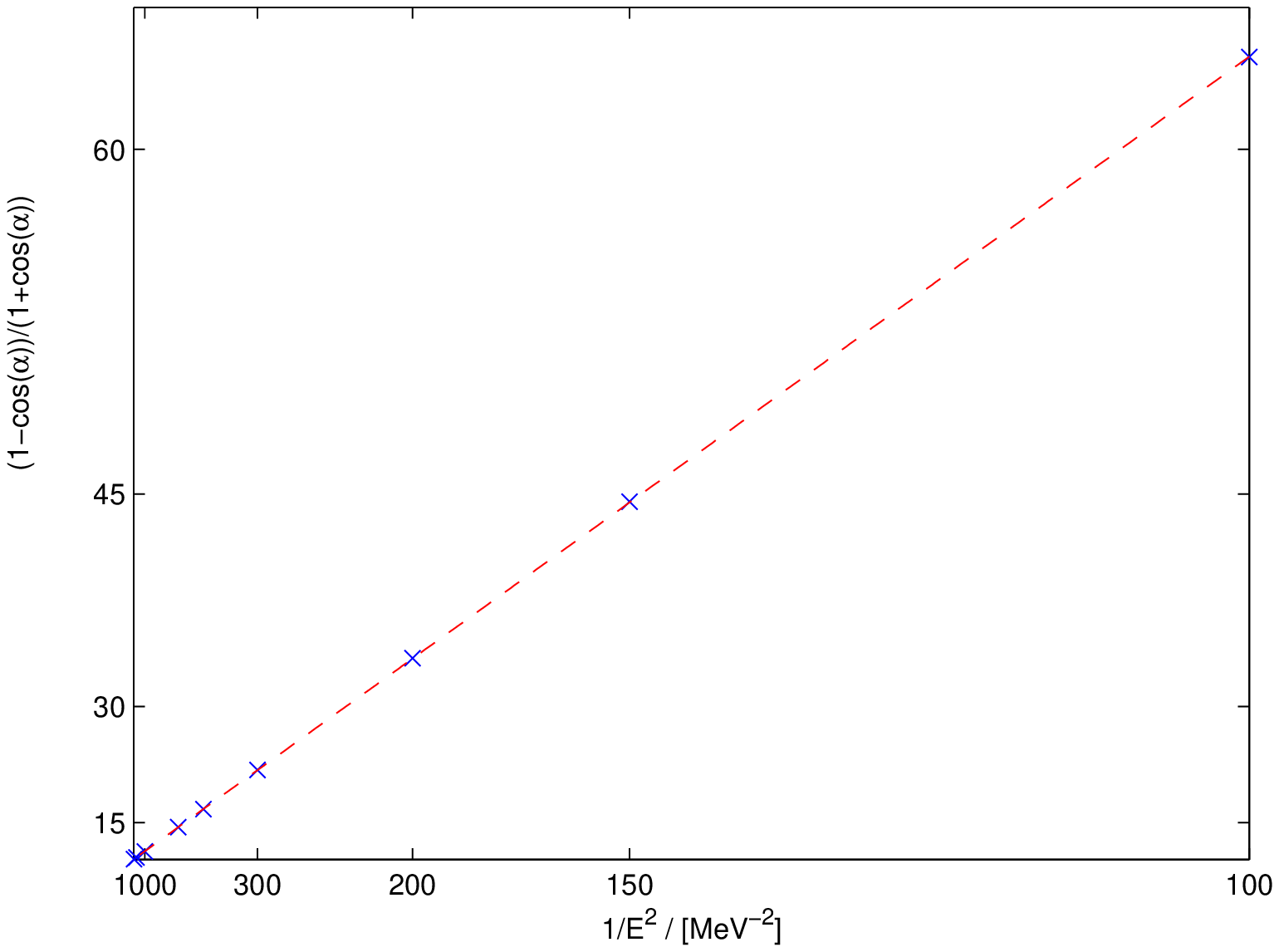}
\caption{ The simulated results (cross) and the analytical calculation (line) are in a good contradiction ($\pm 2\%$). 
\label{fig:anal_vs_simul} }
\end{center}
\end{figurehere}

This enables us to plot the bending angle vs. the electron's energy (Fig. \ref{fig:bending_energy}). For large bending radius %%@
$\rho\,\gg\,R_{magnetic}$ the analytic expression as derived in the appendix can be simplified. 

$ $

$$\begin{array}{c}
\alpha\,=\,arccos\left(\frac{\rho^2-R^2}{\rho^2+R^2}\right)\,=\,arccos\left(\frac{1-\left(\frac{R}{\rho}\right)^2}{1+\left(\frac{R}{\%%@
rho}\right)^2}\right) \\ 
\begin{array}{c}
	   \\
    \longrightarrow \\
    Taylor
  \end{array}
  \,2\,\left(\frac{R}{\rho}\right)\,\propto\,\frac{1}{E}
  \end{array}$$

$ $

This shows that at large energies the bending angle $\alpha$ is inverse proportional to the energy. Therefore the resolution %%@
decreases dramatically with increasing energy. 

$ $

\subsubsection{Simple energy resolution} 

A mono energetic beam with a certain divergence spread passes a drift length and then the spectrometer. During the propagation the %%@
beam diverges and the signal in the detection plane gets spread. For a free drift and if space charge effects are neglected the %%@
spread after a distance $\tilde{Z}$ is given by $\tilde{X}\,=\,Z\,X^{\prime}$. Therefore an increase in divergence results in a %%@
decrease in energy resolution. 
This can be seen by a simple case of a none mono energetic beam. If a beam with a spread in divergence passes the experimental setup, %%@
the beam will be transformed as indicated in Fig. \ref{fig:Entartung}. While the beam passes the dipole magnet electrons with higher %%@
momentum get deflected less than electron with smaller momentum $\left(\,\rho\,=\,\frac{p}{e\,B}\,\right)$. Therefore the electrons %%@
are separated in space according to there momentum. This is indicated in Fig. \ref{fig:Entartung}. 

This allows to calculate the spectrometer's resolution in a simple way. To each point y in the detection plane a certain energy range %%@
$\Delta E$ can be assigned (green line in Fig. \ref{fig:Entartung}). The energy resolution is than defined as $\frac{\Delta E}{E}$. %%@
For a setup with 1 meter drift length between source and magnet edge and a additional drift length of 0.2 meter between magnet edge %%@
and detector the energy resolution vs. the energy can be calculates for different divergence (as given in Fig. %%@
\ref{fig:Resolution_01}). 
Since the spread in y is proportional to the drift length the energy resolution depends crucial on the distance between source and %%@
magnet (Fig. \ref{fig:Resolution_3D_plot}). 

\begin{figurehere}
\begin{center}
\includegraphics[width=7cm]{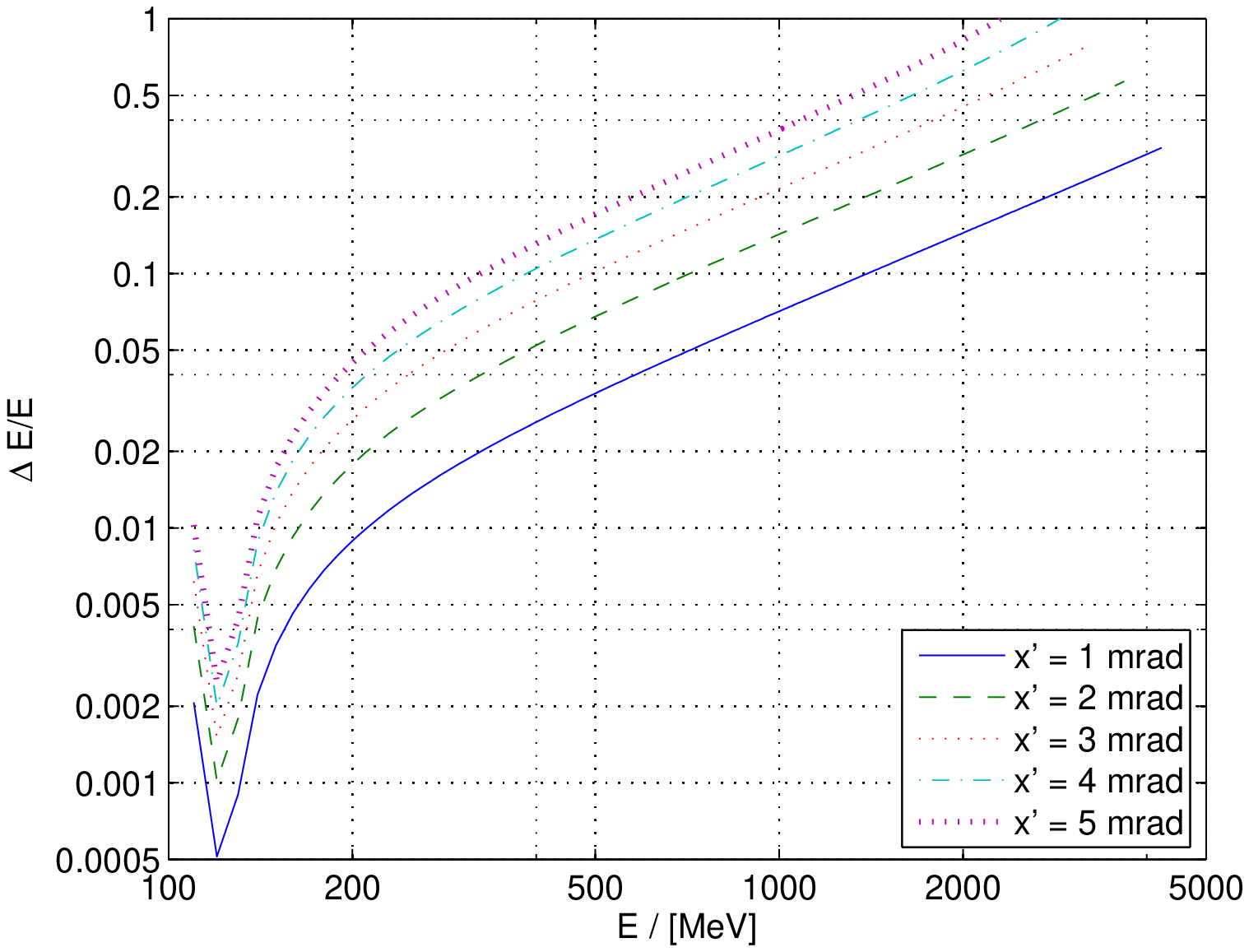}
\caption{ The plot shows the energy resolution $\frac {\Delta E}{E}$ vs. the electron beam's energy E. A focusing effect can be seen %%@
at around 110 MeV. For a setup with 1 meter drift length between source and magnet edge and an additional drift length of 0.2 meter %%@
between the magnet edge and the screen. The magnetic field was the real measured field at $87\,A$ ($1.15\,T$). 
\label{fig:Resolution_01} }
\end{center}
\end{figurehere}

\begin{figurehere}
\begin{center}
\includegraphics[width=7cm]{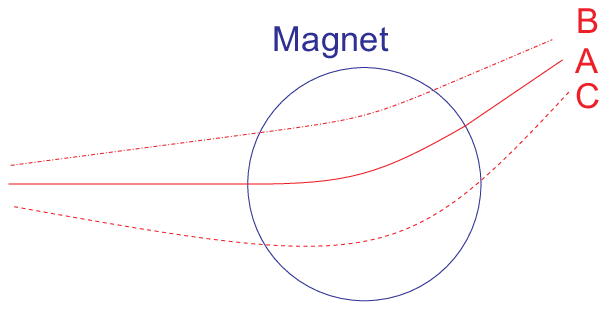}
\caption{ Sketch for three different path. A is a electron going through the magnets centre, B and C cut the magnetic area off centre %%@
due to there finite divergence. It can be seen that B is staying in the magnetic region for the shortest time, where C is in the %%@
magnetic region for the longest time. The difference in path results in a focusing. 
\label{fig:Focusing} }
\end{center}
\end{figurehere}

It can be seen in Fig. \ref{fig:Resolution_01} that at a certain energy a focusing effect occurs. This can be illustrated by looking %%@
on two rays, one of the ray is following the central line A and the other ray is drifting to the left site B, due to its finite %%@
divergence. Since the magnet has circular shaped pole the pathway in the magnet for the ray A is longer than the pathway for B (see %%@
Fig. \ref{fig:Focusing}). Therefore ray A gets more deflected than B. On the other hand ray C which is drifting to the right of the %%@
central ray is staying in the magnetic field for a longer time than A and B and gets therefore more deflected. This is similar to the %%@
effect described in \cite{Banford}. 

\begin{figurehere}
\begin{center}
\includegraphics[width=7cm]{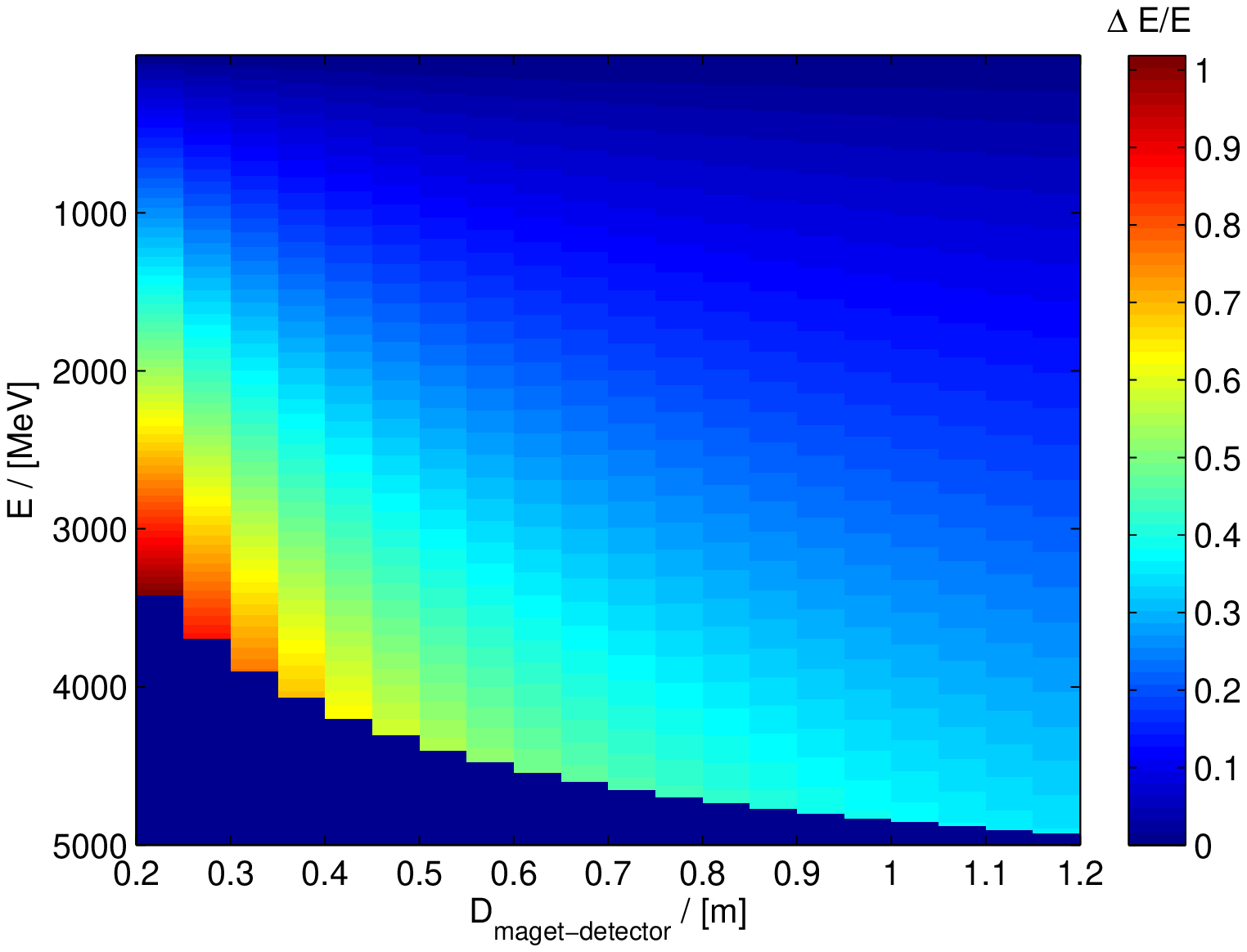}
\caption{ Resolution vs. beam energy and distance between spectrometer and screen. The setup uses a real magnetic field generated by %%@
$87\,A$ coil current and 1 meter drift length between source and magnet edge. 
\label{fig:Resolution_3D_plot} }
\end{center}
\end{figurehere}

\begin{figurehere}
\begin{center}
\includegraphics[width=7cm]{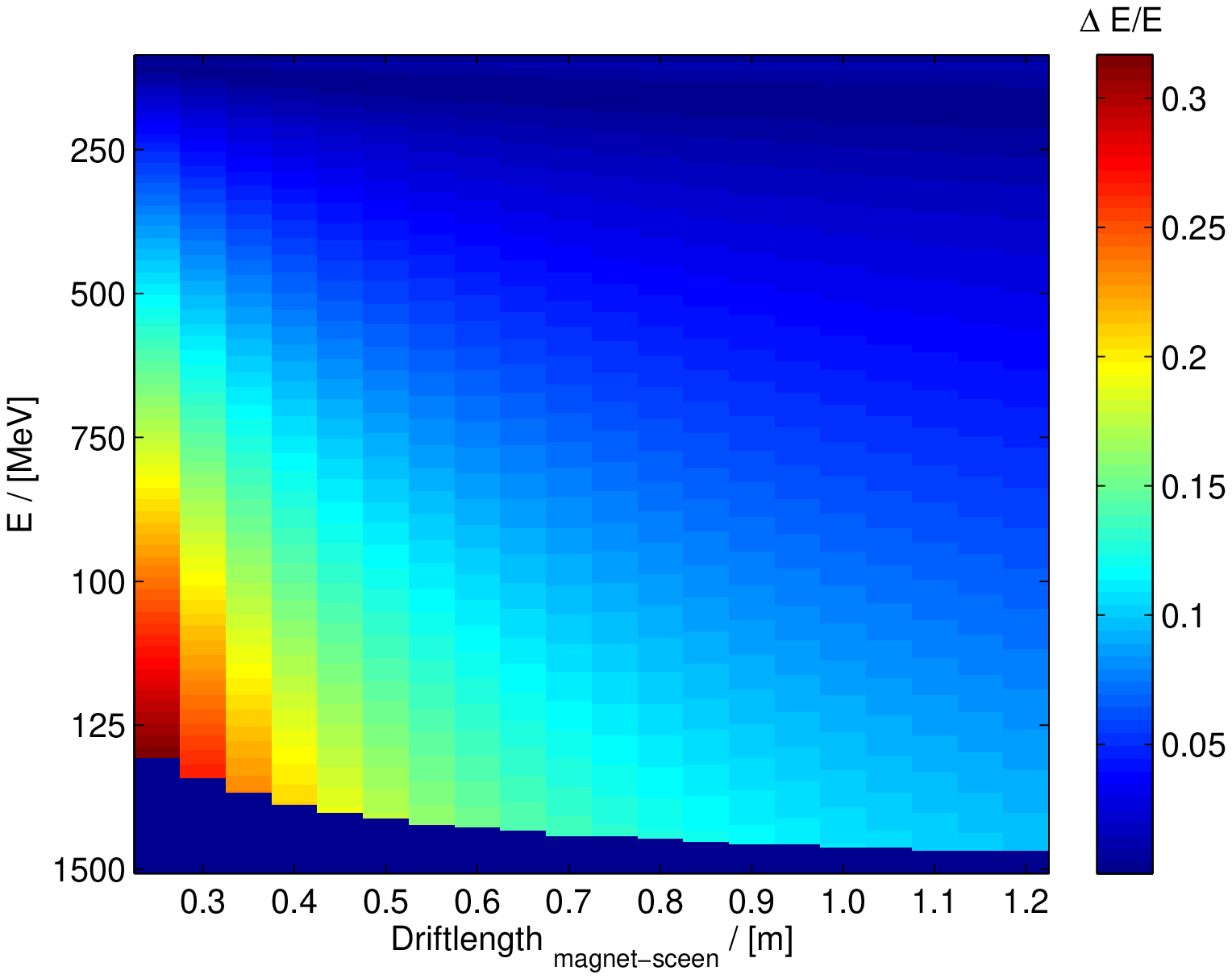}
\caption{ Resolution vs. beam energy and distance between spectrometer and screen. The setup uses a real magnetic field generated by %%@
$87\,A$ coil current and 1 meter drift length between source and magnet edge. 
\label{fig:Resolution_3D_plot_ausschnitt} }
\end{center}
\end{figurehere}

It can be seen in Fig. \ref{fig:Resolution_3D_plot} how the focusing effect is depending on the drift length between magnet and %%@
screen. A focusing is achieved mainly on energies between 100 MeV and 200 MeV. The results shown in Fig. \ref{fig:Resolution_3D_plot} %%@
have been made without taking space charge effects into account \cite{Humphries}. 

$ $

\begin{figurehere}
\begin{center}
\includegraphics[width=7cm]{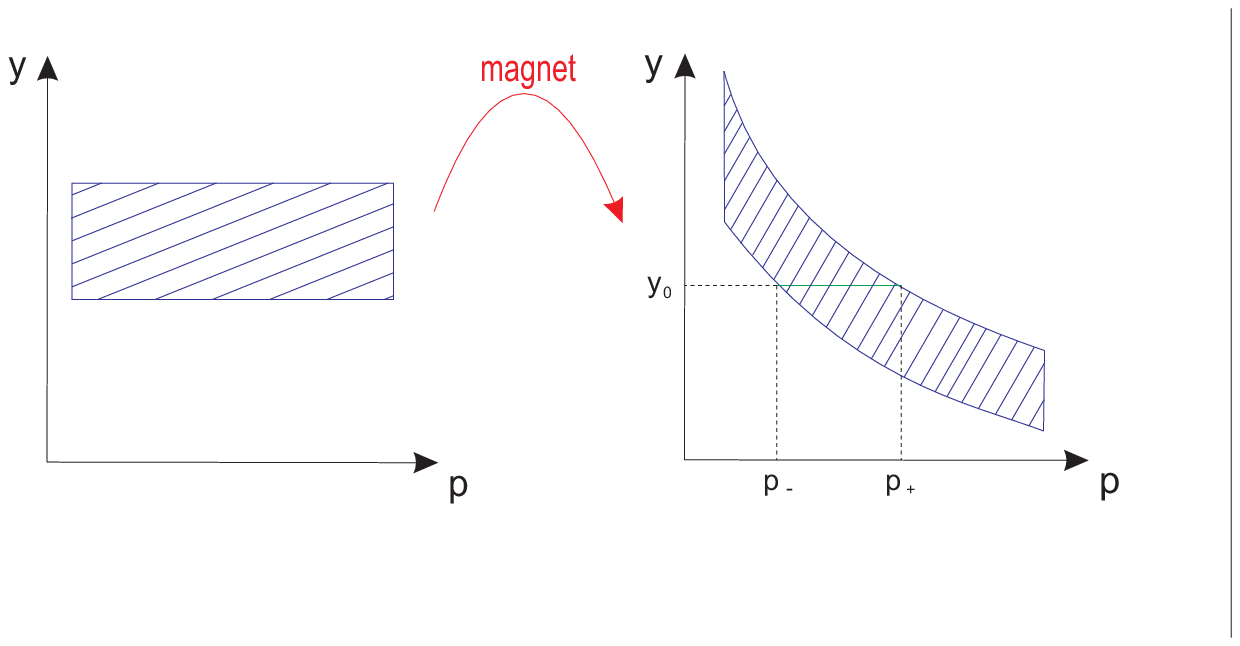}
\caption{ A schematic picture of the evolution of a area in p-y space after passing the spectrometer. 
\label{fig:Entartung} }
\end{center}
\end{figurehere}

\subsubsection{Lifting the degeneracy} 

However if the divergence is known the degeneracy can be lifted. Measuring the divergence can be achieved by putting a screen $T_1$ %%@
in front of the spectrometer (Fig. \ref{fig:Faltung}). This allows to measure the beam's size and position. Now the opportunity is %%@
given to simulate the propagation of different mono energetic electron beam's (at energy $\{E_i\}_{i\in I}$) with the determined %%@
divergence and initial position. Once the propagation of mono energetic electron beam is known the distribution $S(E_i)$ on the %%@
detection screen $T_2$ can be calculated easily. Since it is known how mono energetic beams map on the detection screen $T_2$ as a %%@
function of their energies the opportunity is given to write an arbitrary signal $S$, which is detected on $T_2$, as a linear %%@
combination of mono energetic beams $S(E_i)$. 

$ $

$$\tilde {S}\,=\,\Sigma _{i\in I}\,\lambda _i\,S(E_i) $$ 

$ $

$\tilde {S}$ can be fitted to the measured distribution $S$ by a root mean square method, with the restriction that $\lambda _i \geq %%@
0$ for all $i\in I$. The set $\{\lambda _i\}_{i\in I}$ is than equal the beams energy distribution. 

\begin{figurehere}
\begin{center}
\includegraphics[width=7cm]{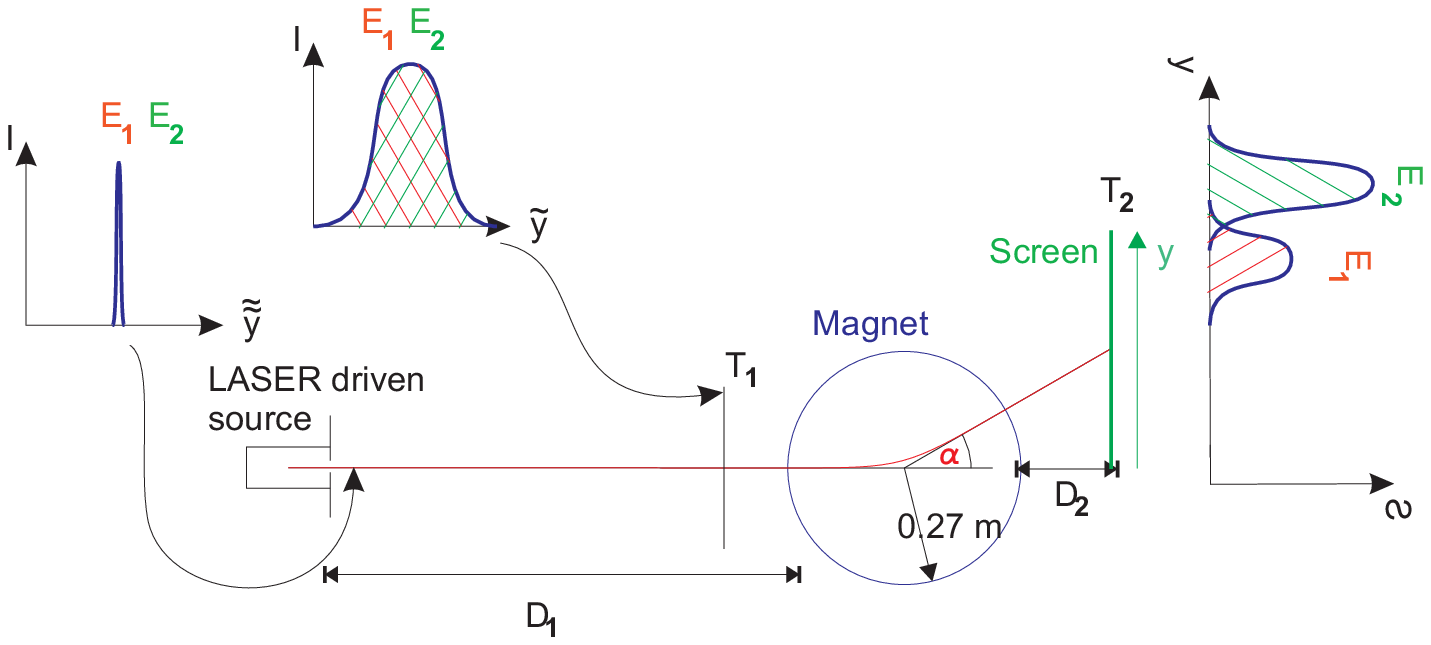}
\caption{ The sketch gives a proposal for an experimental setup to measure the energy resolution more accurate. $T_1$ and $T_2$ %%@
illustrate two screens which measure the beam's position and spatial distribution simultaneously. For simplicity the picture shows a %%@
beam consisting of electrons at two different energies. After a certain drift length $D_1$ the beam has due to its divergence a %%@
certain spatial distribution. After passing the spectrometer the electrons get deflected in a known way according to their energy. %%@
The result measured on the screen is a superposition of electrons from different energies. 
\label{fig:Faltung} }
\end{center}
\end{figurehere}

To get a rough understanding what influence noise at the detection screen $T_2$ on this analyse procedure has, a mono energetic beam %%@
$E$ with a uniform distribution in divergence (from $-3 \,mrad$ to $3 \, mrad$) was studied. The noise was simulated by broadening %%@
the signal in y direction by $0.5\,mm$. The width of the electron energy distribution was determined and uses for calculating the %%@
energy resolution (Fig.~\ref{fig:DeltaE_500um_3mrad}).

\begin{figurehere}
\begin{center}
\includegraphics[width=7cm]{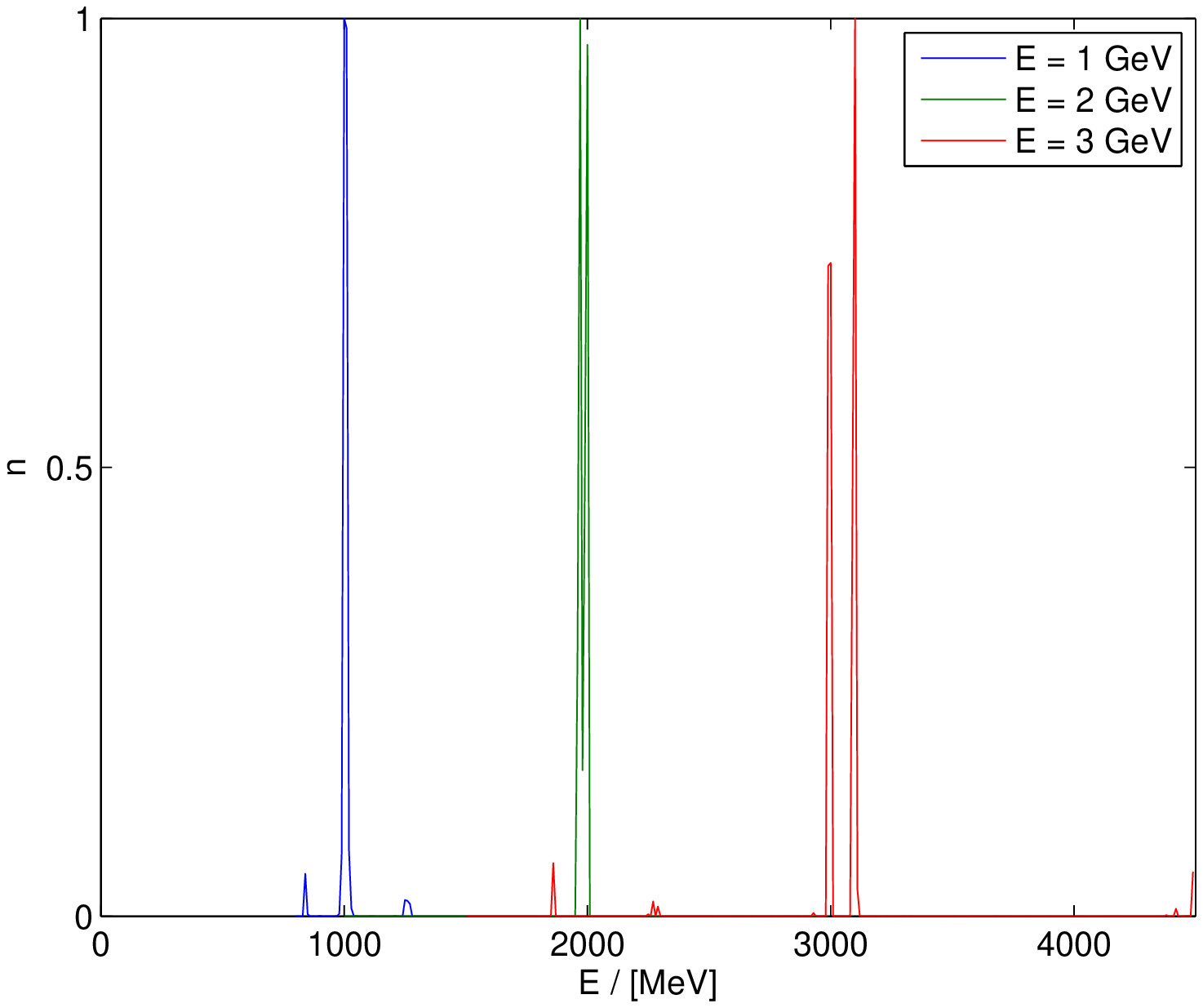}
\caption{ 
\label{fig:DeltaE_500um_3mrad} }
\end{center}
\end{figurehere}

This results are only very rough estimation. The inaccuracy from the screen $T_1$ has not been taken in to account. Also only mono %%@
energetic beams were studied. However this maybe far from real experimental data. 

$ $

\subsubsection{Motion in z direction} 

The bulk field does not have any influence on the motion's z component. Therefore only the stray field accelerates the particle along %%@
the z direction. Due to \cite{Banford} the stray field at the magnet edge causes a defocusing in the z direction. Indeed this can be %%@
observed in the numerical simulation \ref{fig:Beugung_an_edge_traject}. The defocusing as a function of the energy can be simulated %%@
(Fig. \ref{fig:Beugung_an_edge_E}). 

\begin{figurehere}
\begin{center}
\includegraphics[width=7cm]{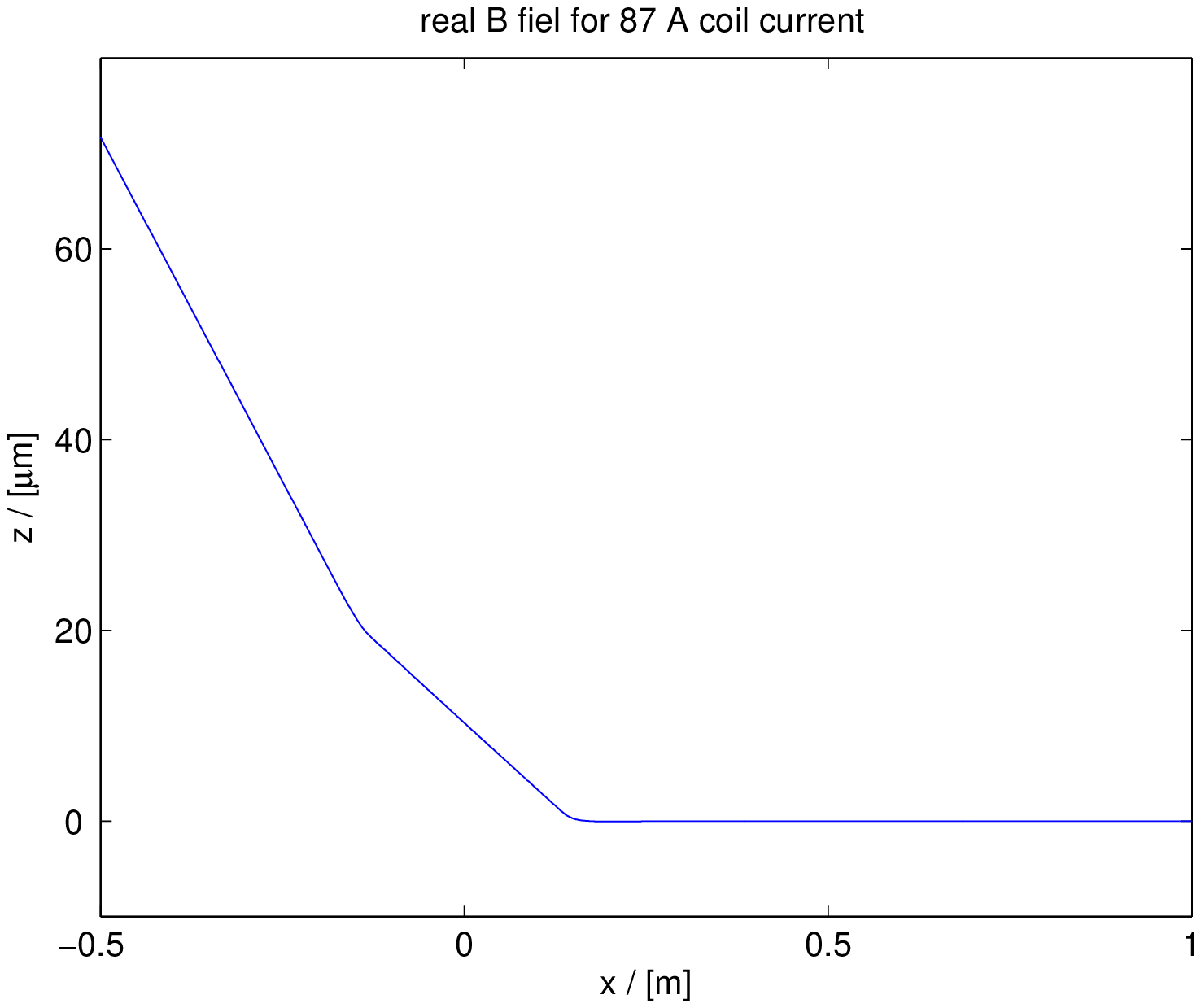}
\caption{ The x component vs. the z component of the trajectory. It can be seen that the electron propagates freely except at the
magnet edge where it gets bended. For the simulation no space charge effect have been included. The simulated electron had none
divergence and a initial position at $\left(1.27,\,-0.001,\,0)\right)\,m$. For the spectrometer the real magnetic field at $87\,A$ %%@
coil current was used. \label{fig:Beugung_an_edge_traject}}
\end{center}
\end{figurehere}

\begin{figurehere}
\begin{center}
\includegraphics[width=7cm]{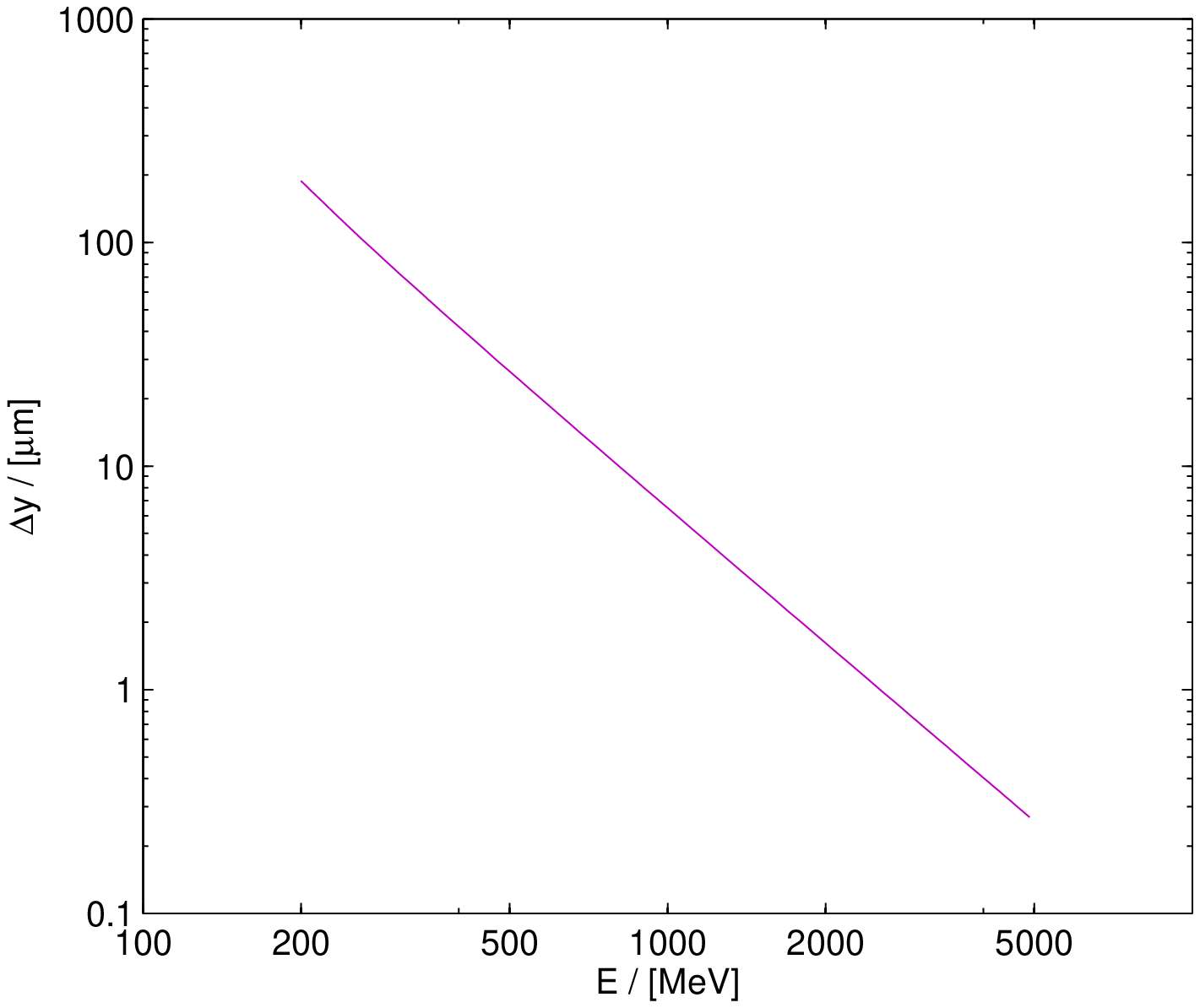}
\caption{ The displacement in z direction yielding from bending effects at the magnet's edge vs. the beam energy. For the simulation %%@
no space charge effect have been included. The simulated electron had none divergence and a initial position at %%@
$\left(1.27,\,-0.001,\,0)\right)\,m$. For the spectrometer the real magnetic field at $87\,A$ coil current was used. 
\label{fig:Beugung_an_edge_E} }
\end{center}
\end{figurehere}

However the broadening created by the magnet edge is small compared with the beam's divergence that can be expected. 

\section{Conclusion} 

For the magnets physical properties it can be stated that the magnetic radius is $(167.0 \pm 0.6)\,mm$. A stray field occurs over a %%@
length of $(110 \pm 5)\,mm$. However the magnetic field's shape $f(r)$ does not depend on the coil current. Furthermore the maximal %%@
magnetic field $B_z$ was measured for different coil current and it was shown that the field is proportional to the current. In %%@
addition a memory effect for the magnetic field could be observed, depending on the path that was chosen to adjust a certain current. %%@
In the adiabatic limes the different between lowering and rising direction was found to be smaller than $0.015\,T$. For further %%@
studies it could be of interest to measure memory effects beyond the adiabatic limit.  

Numerical simulation have shown how the bending angle depends on the electrons energy. It can be concluded that the bending angle %%@
does not change much for large energies ($E\,\geq\,1\,GeV$). This decreases the energy resolution dramatically for increasing %%@
energies. However at low energies the resolution is relatively high ($1^o/_{oo}$ for $E\,=\,150\,MeV$). Since some interests lies in %%@
electron energies up to $5\,GeV$ a proposal has been made to resolve energy spread in this regime energies. However much more studies %%@
have to be done to clarify if this procedure is appropriated or not. 

In addition focusing effects in y direction and defocusing effects in z direction has been studied. It was shown that this effects %%@
are stronger on low energies ($\leq 300\,MeV$).

\section{Appendix} 

If it is assumed that the magnetic field is homogeneous in a circular area with radius $R_{magnetic}$ and zero other wise %%@
($\vec{B}\,=\,B_{max}\,\vec{e}_z\,\chi _{\sqrt{x^2+y^2}\leq R_{magnetic}}$), a analytical solution for the bending angle can be %%@
found. For simplicity it is further assumed that the incident beam is perpendicular to the magnet's surface. Symmetry reasons request %%@
that the  outgoing beam is perpendicular as well. Since the magnetic field in the circular region is constant the electron's %%@
trajectory will be a segment of a circle. The bending radius can easily be calculated by $\rho\,=\,\frac{p}{e\,B}$. Knowing the %%@
bending radius, the magnetic radius and the angle of incident ($\beta\,=\,0$) allows us to calculate the bending angle $\alpha$ (see %%@
Fig.\ref{fig:Bending}). 

$ $

$$\alpha\,=\,arccos(\frac{\rho^2-R^2}{\rho^2+R^2})$$

$ $

The simulated results for the bending angle are plotted on a scale which shows a deviation from the analytical value as a deviation %%@
from the straight line Fig. \ref{fig:anal_vs_simul}. Even though that the analytical method gives quite good results the accuracy is %%@
not high enough. 

\begin{figurehere}
\begin{center}
\includegraphics[width=7cm]{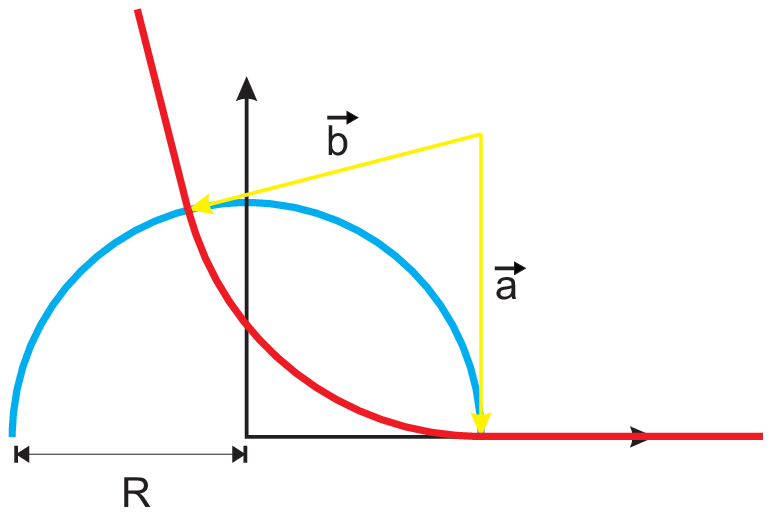}
\caption{ The sketch shows the trajectory of an electron in a magnetic field given by $\vec{B}\,=\,B\,\vec{e}_z\,\chi %%@
_{\sqrt{x^2+y^2}\leq R_{magnetic}}$. Inside the magnetic field the electron follows a circle and outside the electron follows a %%@
straight line. 
\label{fig:Bending} }
\end{center}
\end{figurehere}

\subsection{Acknowledgment} 

On the experimental side the work has been supported by Ron Morton and George Hammett from the electrical work shop, Peter Shrimpton %%@
and Keith Long from the teaching facility and Mick Williams from the technical work shop at the University of Oxford.

}
\end{multicols}

\end{document}